\documentclass[11pt,a4paper]{article}%
\pdfoutput=1
\usepackage{jheppub}
\usepackage{amsmath,amssymb,amsfonts,wasysym}
\usepackage{mathrsfs}
\usepackage[bf,footnotesize]{caption2}
\usepackage{multirow}
\usepackage{color}
\usepackage{slashed}
\usepackage[]{hyperref}%
\hypersetup{
colorlinks=true,
linkcolor=red,
anchorcolor=black,
citecolor=blue,
filecolor=cyan,
menucolor=red,
runcolor=filecolor, 
urlcolor=magenta,
bookmarks=true,
bookmarksnumbered=true
}
\def\l{\label}
\def\La{\mathcal{L}}

\def\({\left(}
\def\){\right)}
\def\<{\langle}
\def\>{\rangle}
\def\f{\frac}
\def\be{\begin{equation}}
\def\ee{\end{equation}}
\def\bry{\begin{array}}
\def\ery{\end{array}}
\def\bes{\begin{subequations}}
\def\ees{\end{subequations}}
\def\bit{\begin{itemize}}
\def\eit{\end{itemize}}
\def\ben{\begin{enumerate}}
\def\een{\end{enumerate}}
\def\dst{\displaystyle}

\def\de{\partial}

\def\ovl{\overline}
\def\Tr{\text{Tr}}

\newcommand{\Dsl}{D\llap{/\kern+1.5pt}}
\newcommand{\dsl}{\partial\llap{/\kern+0.5pt}}
\newcommand{\psl}{p\llap{/\kern-0.5pt}} 
\newcommand{\uno}{{\scriptscriptstyle{{\widehat {\mathbf1}}}}}

\newcommand{\Eqr}[1]{Eq.~\eqref{#1}}

\newcommand{\warp}{\log\tfrac{z_{IR}}{z_{UV}}}

\definecolor{grey}{rgb}{0.6,0.6,0.6}
\newcommand{\red}[1]{{\color{magenta}\color{red}#1\color{magenta}}}
\newcommand{\blue}[1]{{\color{blue}\color{blue}#1\color{blue}}}

\newcommand{\grey}[1]{{\color{grey}\color{grey}#1\color{grey}}}

\title{A minimally tuned composite Higgs model from an extra dimension}
\author[a,b]{Duccio Pappadopulo}
\author[c]{Andrea Thamm}
\author[d, e]{Riccardo Torre}
\affiliation[a]{Department of Physics, University of California, Berkeley, CA 94720, USA}
\affiliation[b]{Theoretical Physics Group, Lawrence Berkeley National Laboratory, Berkeley, CA 94720, USA}
\affiliation[c]{Institut de Th\'eorie des Ph\'enom\`enes Physiques, EPFL,  CH--1015 Lausanne, Switzerland}
\affiliation[d]{Dipartimento di Fisica e Astronomia, Universit\`a di Padova, and \\ INFN Sezione di Padova, Via Marzolo 8, I-35131 Padova, Italy}
\affiliation[e]{SISSA, Via Bonomea 265, I-34136 Trieste, Italy}
\emailAdd{pappadopulo@berkeley.edu}
\emailAdd{andrea.thamm@epfl.ch}
\emailAdd{riccardo.torre@pd.infn.it}

\proceeding{\vspace{-1.65cm}\flushright{\small
DFPD-2013/TH/04\\
LPN13-022\\}}
\abstract{
We construct and study the 5D realization of a composite Higgs model with minimal tuning. The Higgs is a (pseudo-)Goldstone boson from the spontaneous breaking of a global $SO(5)$ symmetry to an $SO(4)$ subgroup. The peculiarity of our construction lies in the specific choice of the $SO(5)$ representations of the 5D fermions from which the Standard Model fields arise as chiral zero modes. This choice reduces the tuning of these models to the minimal model-independent value allowed by electroweak precision tests. We analyse the main differences between our 5D construction and other descriptions in terms of purely 4D field theories. 5D models are generally more constrained and show a generic difficulty in accommodating a light Higgs without reintroducing large corrections to the $\hat S$ parameter. We propose a specific construction in which this tension can be, even though accidentally, relaxed. We discuss the spectrum of the top partners in the viable regions of parameter space and predict the existence of light exotic quarks, $\Upsilon$, of charge 8/3 whose striking decay channel $\Upsilon\to W^+W^+W^+ b$ can lead to either exclusion or confirmation of the model in the near future.
}
\begin{document} 
\maketitle
\setcounter{page}{2}


\section{Introduction}

A satisfactory understanding of ElectroWeak Symmetry Breaking (EWSB) requires the existence of new physics at a scale, $\Lambda_{\text{NP}}$, not far above the weak scale. We expect the new dynamics at $\Lambda_{\text{NP}}$ to provide a solution to the hierarchy problem affecting the Higgs sector in the Standard Model (SM).

Constraints on generic new physics require $\Lambda_{\text{NP}}$ to be at or above the TeV scale. Naturalness, on the other hand, demands low scale new physics. A completely natural theory would, for instance, require the presence of a new set of particles and interactions at a scale
\be\label{topnaturalness}
\Lambda_{\text{NP}}\lesssim\frac{2\pi}{\sqrt 3 y_t} m_h\approx 450\, {\text{GeV}},
\ee
in order to cancel the quadratically divergent top quark contribution to the SM Higgs potential.

The two main frameworks in which the hierarchy problem is usually addressed are supersymmetry and compositeness. The supersymmetric way has been and still is a theorist's favorite. This rank is well deserved by the theoretical appeal of supersymmetry and its ability to make definite predictions due to its weakly coupled nature. Compositeness remains a much less studied alternative.

In the past decade a realistic framework has emerged \cite{Contino:2003p378,Agashe:2004ib,Agashe:2005p372,Contino:2006fd,Panico:2005kd,Barbieri:2007bh,Contino:2010mr} in which the Higgs boson arises as a pseudo-Goldstone Boson (pGB)  
from the spontaneous breaking of a global symmetry $G$, of a new strongly interacting sector, to a subgroup $H$.
These models have two crucial advantages over plain technicolor models. Firstly, the presence of a light Higgs boson allows the parametric separation between the $G\to H$ breaking scale $f$ and the electroweak symmetry breaking scale $v$  which is needed to alleviate the tension of technicolor models with electroweak precision tests \cite{Barbieri:2012wia}.\footnote{A fortuitously light scalar can also be present in technicolor models. For instance it can be the dilaton.} Secondly, the flavor problem of technicolor can be greatly improved by the implementation of partial compositeness \cite{1991NuPhB.365..259K}. This implies that those degrees of freedom, whose elementary nature is well probed (the light quarks and leptons and the transverse polarizations of the gauge bosons), are external to the strongly interacting sector and they communicate with it only through linear couplings of the form
\be\label{pc}
\La_{\text{mix}}= g\, A_\mu \mathcal J^\mu +\lambda\, q\,\mathcal O+\ldots \, ,
\ee 
where $\mathcal J$ is a global current of the strong sector which is  gauged by the SM vector bosons $A_\mu$ and $\mathcal O$ is a fermionic operator (assumed to have a naive mass dimension $5/2$) \cite{KerenZur:2012td}.\footnote{\Eqr{pc} has to be understood as evaluated at the typical mass scale $m_\rho$ characterizing the resonances of the strong sector. If, above this scale, the strong sector behaves as an approximately conformal field theory up to energies of order $\Lambda_{\text{UV}}$, one expects $\lambda(m_\rho)\sim \lambda(\Lambda_{\text{UV}})\left(\frac{m_\rho}{\Lambda_{\text{UV}}}\right)^{d_{\mathcal O}-5/2}$ where $d_{\mathcal O}$ is the scaling dimension of $\mathcal O$ at the fixed point.}

The couplings of the elementary fields to the strong sector break the global symmetry $G$ explicitly and generate a potential for the Higgs which is calculable in specific models and whose general structure we will sketch following the discussion in Ref.~\cite{Panico:2012vr} closely. The largest contribution to the pGB potential comes typically from the interactions generating the top quark mass. The Yukawa for the top quark arises from the coupling of the elementary $q_L$ and $t_R$ to the strong sector according to the pattern sketched above in \Eqr{pc}
\be\label{pctop}
\mathcal L_{\text{mix}}=\lambda_L f\, q_L  \mathcal  O^q_L+\lambda_R f\, t_R \mathcal O^t_R+{\textrm{h.c.}}\,,
\ee
where $f$ is the sigma model decay constant and $\mathcal O$ stands for fermionic resonances of mass $m_\psi$ in the low energy theory (their naive dimension is now $3/2$). Dimensional analysis ensures the leading contribution to the top Yukawa to be
\be
y_t\sim \frac{\lambda_L \lambda_R}{g_\psi}\,,
\ee
where $1<g_\psi\sim m_\psi/f<4	\pi$ is the typical coupling among these fermionic resonances.\footnote{We wish to distinguish this coupling from the one among the vector resonances, $g_\rho$. Notice that $g_{\psi}$ can be naturally smaller than $g_\rho$ due to an approximate chiral symmetry.}
%
%
The dimensionless parameters $\epsilon_{L,R}\equiv \lambda_{L,R}/g_{\psi}$ determine the degree of compositeness of the various SM fields ranging from 0 (elementary state) to 1 (composite state).
 
Deep insight into the structure of the Higgs potential is obtained by exploiting the symmetry properties of \Eqr{pctop}. As the strong sector obeys a global symmetry $G$ the operators $\mathcal  O^q$ and $\mathcal O^t$ can be classified according to their transformation properties under $G$.\footnote{The fact that $G$ is spontaneously broken to a subgroup $H$ does not affect this statement as the Goldstone bosons from $G\to H$ can be used in the definition of $\mathcal O^q$ and $\mathcal O^t$ in order to make them transform linearly under $G$.}
Furthermore when both interactions in \Eqr{pctop} and the SM gauge interactions are switched off, the action for the elementary fields is invariant under an independent \mbox{$SU(2)\times U(1)$} group, with charges corresponding  to their SM quantum numbers. The couplings $\lambda_L$ and $\lambda_R$ break this large $\mathcal G\equiv SU(2)\times U(1)\times G$ to the gauged $SU(2)_L\times U(1)_Y$ vector subgroup. It is however possible to assign spurious transformation properties to the $\lambda$s, promoting them to $\hat \lambda$s, in order to make \Eqr{pctop} formally invariant under $\mathcal G$
\be\label{pctopinv}
\mathcal L_{\text{mix}}= f\, q_L^\alpha (\hat \lambda_{L})_I^\alpha \mathcal O^{q\,I}_L + f\, t_R^\alpha (\hat \lambda_{R})^\alpha_I \mathcal O^{t\,I}_R+{\textrm{h.c.}}\,;
\ee
$\alpha$ and $I$ are irreducible $SU(2)\times U(1)$ and $G$ indices respectively. 

When both the strong sector and the high energy fluctuations of the elementary modes are integrated out the resulting effective action will respect the spurious $\mathcal G$ symmetry. $SU(2)\times U(1)$ invariance implies that, for instance, at leading order in $\lambda_L$ only the combination $X_{L\,I\otimes J}\equiv (\hat \lambda_L \hat \lambda_L)_{I\otimes J}$ will enter the Higgs potential. The same argument constrains the form of the $\lambda_R$ contribution. 

$X_{L,R}$ transform spuriously as reducible representations of $G$: $X_{I\otimes J}=\sum_r X^r_{I_r}$. In order to build non-trivial (Higgs dependent) invariants that can contribute to the Higgs potential we introduce the Goldstone boson matrix $U(h)$ transforming as \cite{Callan:1969p1799} 
\be\label{goldstones2}
U(h)\to g\, U(h) \, \hat h(g,h)^{-1}, \qquad g \in G,~\hat h \in H.
\ee
$U(h)$ can be used to turn irreducible representations of $G$ into reducible representations of $H$. Combining $X$ with $U$ according to their transformation properties we can construct non-linear $G$ invariants, $I_i(h/f)$, which depend non-trivially on the pGB $h$.
The structure of the potential will thus be \cite{Giudice:1024017} 
\be\label{potentialloops}
\bry{lll}
\dst V(h)&=&\dst V^{({\textrm{1\,loop}})}(h/f)+V^{({\textrm{2\,loop}})}(h/f)+\ldots  \vspace{2mm}\\
&=&\dst f^2m_{\Psi}^2\left(\frac{g_{\psi}}{4\pi}\right)^2\left(\epsilon^2 \mathcal F_1^{(1)}(h/f)+\epsilon^4 \mathcal F_2^{(1)}(h/f)+\ldots\right)\vspace{2mm}\\
&&+\dst f^2m_{\Psi}^2\left(\frac{g_{\psi}}{4\pi}\right)^4\left(\epsilon^2 \mathcal F_1^{(2)}(h/f)+\ldots\right)+\ldots\,,
\ery
\ee
where $f$ is the decay constant of the pGB, $h$ is the physical Higgs field and 
\be
\mathcal F=\sum_{i} c_{i} I_{i}\(\f{h}{f}\),
\ee
i.e.~each of the functions $\mathcal F$ is a sum of the non-trivial $G$ invariants which can be constructed at a given order in $\epsilon$ and $\hbar$. \Eqr{potentialloops} allows us to determine how EWSB occurs. Generically we expect the one-loop, leading order $\mathcal F_1^{(1)}$ contribution to be dominant. As the various $\mathcal F$ are expected to be $O(1)$ functions of their argument, some degree of cancellation among the $c_i$ is necessary to obtain a small enough ratio $\xi=(v/f)^2$ which is crucial for a realistic phenomenology. In the absence of other cancellations, this irreducible tuning of a composite Higgs model can be quantified by $\xi$ itself.

The simplest realistic realization of the composite Higgs idea is represented by \sloppy\mbox{$G=SO(5)\times U(1)_X$} and $H=SO(4)\times U(1)_X$.\footnote{The $U(1)_X$ factor is needed to obtain the correct hypercharge for the SM fermions.} The coset space $G/H$ contains a quadruplet of GB, three of which are eaten by the SM gauge bosons while the fourth is the physical Higgs boson. Since the Higgs is an angular variable on the compact coset manifold $G/H\sim S^4$, the $G$ invariants are expected to be trigonometric functions of $h/f$. 

For certain choices of the operators $\mathcal O^{q,t}$, the leading contribution to the Higgs potential, $\mathcal F^{(1)}_1$, contains a single invariant at $O(\epsilon^2)$.
This happens for instance in the simplest case in which both $\mathcal O^{q}$ and $\mathcal O^{t}$ transform as a $\mathbf 5$ of $SO(5)$ where
\be\l{leadingorderterm}
\epsilon^{2}\mathcal F^{(1)}_{1}=c_{1}\epsilon^{2} s_{h}^{2}\,,
\ee
with $s_h\equiv \sin h/f$, and only a discrete set of values is available for $\xi$. In order to be able to tune $\xi\ll 1$, it is necessary to suppress the coefficient of the leading order term in \Eqr{leadingorderterm} to be of the same order as the $O(\epsilon^4)$ subleading contribution which will contribute with the new independent functions of $h/f$
\be
\epsilon^{4}\mathcal F^{(1)}_{2}=\(c_{2}\epsilon^{2}\)\epsilon^{2} s_{h}^{2}\(1-s_{h}^{2}\).
\ee
In other words we need $c_{1}\approx c_{2}\epsilon^{2}$ and the tuning is worsened to $\xi\times\epsilon^2$.

The simplest situation that avoids this peculiar behavior corresponds to the $\mathcal O^q$ being the $\mathbf{14}$ dimensional (symmetric and traceless) representation of $SO(5)$ and $\mathcal O^t$ being a singlet. This choice defines the MCHM$_{14}$ \cite{Pomarol:2012vn}. In the MCHM$_{14}$, already $\mathcal F^{(1)}_1$ contains two invariants and a viable EWSB can be achieved with the minimal amount of tuning. 

To study the consequences of this model beyond simple dimensional analysis the calculability of the Higgs potential is a crucial property.
A 4D realization of the MCHM$_{14}$ has been obtained in Ref.~\cite{Pomarol:2012vn} using Weinberg's sum rules to ensure calculability.\footnote{For an extensive discussion of the use of generalized Weinberg's sum rules in general composite Higgs models see Ref.~\cite{Marzocca:2012tt}.} Collective breaking, used in Ref.~\cite{Panico:2012vr}, led to the same result. In this paper, we follow the original approach of Ref.~\cite{Agashe:2004ib} studying a holographic realization of the MCHM$_{14}$ where calculability is related to locality in an extra dimension. The only difference between our approach and the original holographic approach to composite Higgs models is the 5D metric, that we choose to be flat and not AdS$_5$.

The paper is organized as follows. In Section~\ref{Sec:spurionanalysis}, we clarify the role of symmetries and dimensional analysis in the structure of the Higgs potential of the MCHM$_{14}$.
We recall the general features of holographic composite Higgs models in Section~\ref{Sec:5Dmodel} and describe the setup leading to the MCHM$_{14}$ in detail.
In Section~\ref{Sec:results}, we compute the Higgs potential and study the dependence of $\xi$ and $m_h$ on the parameters of the 5D model. Here we study the spectrum of the fermionic resonances, their relations with the Higgs mass and the tuning of the holographic implementation.
In Section~\ref{Sec:conclusion} we summarize and conclude, sketching some possible phenomenological implications of the model.


\section{A composite Higgs with $q_L\in {\bf{14}}$ and $t_R\in {\bf{1}}$: spurionic analysis}\l{Sec:spurionanalysis}

In this section we describe some features of the MCHM$_{14}$ following from its symmetries. We begin by discussing the gauge/Goldstone sector and move on to the fermionic sector.

\subsection{The Goldstone and gauge sector}

The bosonic sector of the theory comprises the Goldstone bosons of the $SO(5)/SO(4)$ coset and the gauge bosons of the $SU(2)_L\times U(1)_Y$ SM gauge symmetry. The first appear through the matrix $U(h)$ defined by
\be\label{GB}
U(h)=\exp\left(i\frac{\sqrt{2} h^{\hat a} T^{\hat a}}{f}\right)\,,
\ee
where $T^{\hat a}$ are the broken $SO(5)/SO(4)$ generators (see Appendix~\ref{SO5generators} for their explicit expressions). Notice that $U(h)$ is an orthogonal matrix transforming as in \Eqr{goldstones2}. Moreover, we can use $U(h)$ to construct the 5-vector
\be
\Sigma_I(h)=U_{I5}(h)=\f{1}{h}\sin\f{h}{f}\(h^{1}, h^{2}, h^{3}, h^{4}, h\cot\f{h}{f} \)^T,\qquad h =\sqrt{(h^{i})^{2}}\,,
\ee
transforming linearly under $SO(5)$, i.e.~$\Sigma( h)\to g\,\Sigma(h)$. In the unitary gauge this reduces to 
\be
\Sigma=\left(0,0,0,\sin\frac{h}{f},\cos\frac{h}{f}\right)^T.
\ee
The gauging of the SM $SU(2)_{L}\times U(1)_{Y}$ symmetry is introduced through the usual covariant derivative
\be
D_\mu \Sigma\equiv \partial_\mu \Sigma-i T_L^I W_\mu^I\Sigma-i T_R^3 B_\mu \Sigma.
\ee
For fields with a nonvanishing $X$-charge the hypercharge is given by $Y\equiv T_R^3+X$. In a derivative expansion the leading term in the Lagrangian of the pGB is thus
\be\l{sigmamodel}
\mathcal{L}_{\text{kin}} = \frac{f^2}{2} \left(D_\mu \Sigma\right)^T D^\mu \Sigma\, ,
\ee
such that $m_{W} = g f/2 \sin (\< h \>/f)$. Furthermore, we expect the elementary gauge bosons to couple linearly to dimension 3 conserved currents of the strong sector, as discussed below \Eqr{pc}. Upon integrating out the heavy degrees of freedom of the composite sector these $G$-breaking interactions will generate an effective Lagrangian for the gauge bosons which can be used to derive the gauge contribution to the Higgs potential. 
The most general effective action for the SM gauge fields can be derived under the assumption that the full global symmetry of the strong sector $G$ is gauged. At the quadratic level in momentum space, this $SO(5)\times U(1)_{X}$ invariant Lagrangian is given by
\be\label{Leffgauge1}
\La_{\text{eff}}^{g}=\frac{1}{2}P^{(t)}_{\mu\nu}\left[\Pi_0^X(p) A^{X\,\mu} A^{X\,\nu}+\Pi_0(p)\Tr[A^\mu A^\nu]+\Pi_1(p) \Sigma^T A^\mu A^\nu \Sigma\right] \, ,
\ee
where $P_{\mu\nu}^{(t)}=\eta_{\mu\nu}-p_\mu p_\nu/p^2$ is the transverse projector.\footnote{Notice that $\Sigma^TA\Sigma=0$.}
We identify the SM gauge bosons among the $SO(5)\times U(1)_{X}$ ones according to
\be\label{gaugefields}
A^{a_L}_\mu=W^a_\mu, \qquad A^{3_R}_\mu=A^X_\mu=B_\mu, \qquad A^{1_R,2_R}_\mu=A^{\hat a}_\mu=0\,.
\ee
Using these relations, the effective Lagrangian for the SM gauge fields becomes
\be\l{Leffgauge2}
\La_{\text{eff}}^{g}= P_{\mu\nu}^{(t)}\left[\frac{1}{2}W^{a\mu}\Pi_{ab}W^{b\nu}+W^{3\mu}\Pi_{30}B^{\nu}+\frac{1}{2}B^{\mu}\Pi_{00}B^{\nu}\right]\,,
\ee
where the form factors are related to those in \Eqr{Leffgauge1} by
\be\label{gaugeFF}
\bry{lll}
\Pi_{00}&=&\dst \Pi_0+\Pi_0^X+\frac{s_h^2}{4}\Pi_1\,,\vspace{2mm}\\
\Pi_{03}&=&\dst -\frac{s_h^2}{4}\Pi_1\,,\vspace{2mm}\\
\Pi_{ab}&=&\dst \delta_{ab}\left(\Pi_0+\frac{s_h^2}{4}\Pi_1\right)\,.
\ery
\ee


\subsection{The fermionic sector}

We assume the existence of fermionic operators $\mathcal{O}^{q}_L\in \mathbf{14}_{2/3}$ and $\mathcal{O}^{t}_R\in \mathbf{1}_{2/3}$. The subscript refers to the $U(1)_X$ charge.\footnote{Notice that our choice of the representations of the composite operators mixing with the top sector do not generate dangerous corrections to the $Zb_{L}b_{L}$ coupling thanks to a $P_{LR}$ symmetry \cite{Agashe:2006at,Mrazek:2011iu}. In this paper we do not discuss the implementation of SM fermion masses apart from the top quark since their effect on the Higgs potential is subleading. However, the model can be extended in order to generate a mass for the $b$-quark without introducing large corrections to the $Zb_{L}b_{L}$ and $Zb_{R}b_{R}$ couplings. These will be generically proportional to the mixing angles in the bottom sector which can be small consistently with the generation of the small $b$ mass.} According to partial compositeness these operators couple to the elementary fermions through linear mixings as discussed in the Introduction
\be\label{pctopinvSO5}
\mathcal L_{\text{mix}}= \lambda_{L} f\, q_L^\alpha \Tr \left[ P_{L}^\alpha \mathcal O^{q}_L \right] + \lambda_{R} f\, t_R^\alpha P_{R}^\alpha \mathcal O^{t}_R+{\textrm{h.c.}}\,.
\ee
To make contact with \Eqr{pctopinv}, we define $\hat \lambda_{L} = \lambda_{L} P_{L}$ and analogously $\hat \lambda_{R} = \lambda_{R} P_{R}$ where $\lambda_{L}$ and $\lambda_{R}$ are numerical coefficients describing the mixing strength between the elementary states and the composite sector. Both $P^\alpha_L$  and $\mathcal O_q$ are taken to be $5\times 5$ matrices transforming as
\be
\mathcal O_q\to g^{-1}\mathcal O_q ~g, \quad P_L^\alpha\to g^{-1}P_L^\alpha ~g, \quad g\in SO(5).
\ee
The index $\alpha$ refers to the elementary $SU(2)$ factor discussed in the Introduction. \Eqr{pctopinvSO5} is formally invariant under a spurious symmetry group $\mathcal G\equiv SU(2)\times U(1)\times SO(5)\times U(1)_X$ if one assumes
\begin{equation}
\bry{lll}
q_L\equiv(\mathbf 2, 2/3, \mathbf 1,0), \quad &\mathcal O_q\equiv (\mathbf 1, 0,\mathbf {14}, 2/3),\quad &P_L\equiv (\mathbf 2,-2/3, \mathbf {14}, -2/3)\,, \vspace{2mm} \\
t_R\equiv(\mathbf 1, 2/3, \mathbf 1,0), \quad & \mathcal O_t\equiv (\mathbf 1, 0,\mathbf 1, 2/3),\quad &P_R\equiv (\mathbf 1,-2/3, \mathbf 1, -2/3)\,.
\ery
\end{equation}
The spurion $P^{\alpha}_L$ is given by
\be\label{P}
P_L^{\alpha}= \frac{1}{2}\left(\begin{array}{ccc|c}
~~~&~&~~~& ~\\
~~~&~&~~~& \vec v^{\alpha}\\
~~~&~&~~~& ~\\
\hline
~~~&\vec v^{\alpha\,T}&~~~& ~
\end{array}\right) \, ,
\ee
with
\be
 \vec v^{1\,T}=(0,0,i,-1),\qquad \vec v^{2\,T}=(i,-1,0,0),
\ee
while $P_R=1$. The explicit expression for the matrix $\psi_q\equiv q_L^\alpha P_L^\alpha$ is thus
\be\label{qtimesP}
\psi_q= \frac{1}{2}\left(\begin{array}{cccc|c}
~&~&~&~& i b_L\\
~&~&~&~& -b_L\\
~&~&~&~& i t_L\\
~&~&~&~& - t_L\\
\hline
i b_L&-b_L&i t_L&-t_L& ~
\end{array}\right) \, .
\ee
Once the strong sector is integrated out the resulting effective Lagrangian for the light degrees of freedom has to respect the full spurionic symmetry which implies that $q_L$ and $t_R$ will only enter through the combinations $\psi_q$ and $\psi_t\equiv  t_R$. The relevant terms in the most general effective Lagrangian are thus
\be\l{efflag14}
\begin{array}{lll}
\La_{\text{eff}}&=&\dst \Pi_{0}^{q}\Tr\left[\ovl{\psi}_{q}\,\psl\, \psi_{q} \right]+\Pi_{0}^{t}\ovl{\psi}_{t}\,\psl\, \psi_{t}\vspace{2mm}\\
&&\dst +4\Pi_{1}^{q}\Sigma^{T}\ovl{\psi}_{q}\,\psl\,\psi_{q}\Sigma+\Pi_{2}^{q}\(\Sigma^{T}\ovl{\psi}_{q}\Sigma\)\,\psl\,\(\Sigma^{T}\psi_{q}\Sigma\)\vspace{2mm}\\
&&+M_{1}^{t}\bar{\psi}_{t}\Sigma^{T}\psi_{q}\Sigma+\text{h.c.}\,,
\end{array}
\ee
where $\Pi_{0}^{q},\Pi_{0}^{t},\Pi_{1}^{q},\Pi_{2}^{q},M_{1}^{t}$ are $p^2$-dependent form factors that we will determine in the complete 5D theory in Section \ref{Sec:5Dmodel} (the explicit form in 5D can be found in Appendix \ref{App5Dfermions}). In terms of $t_L$, $b_L$ and $t_R$ the Lagrangian in \Eqr{efflag14} becomes
\be\l{efflag15}
\La_{\text{eff}}^{f}=\dst \Pi^{b_L}\ovl{b}_{L}\,\psl\, b_{L}+\Pi^{t_L}\ovl{t}_{L}\,\psl\, t_{L}+\Pi^{t_R}\ovl{t}_{R}\,\psl\, t_{R}-\(\Pi^{t_L t_R}\ovl{t}_{L} t_{R}+\text{h.c.}\) \, ,
\ee
where the form factors are related by
\be\label{fffermions}
\begin{array}{llll}
\Pi^{b_L}&=&\Pi_0^{q}+2\Pi_{1}^{q}c_{h}^{2},\\
\Pi^{t_L}&=&\Pi_{0}^{q}+\Pi_{1}^{q}\(1+c_{h}^{2}\)+\Pi_{2}^{q}s_{h}^{2}c_{h}^{2},\\
\Pi^{t_R}&=&\Pi_{0}^{t},\\
\Pi^{t_L t_R}&=& M_1^t s_h c_h,
\end{array}
\ee
with $s_h\equiv \sin h/f$ and $c_h\equiv \cos h/f$. \\
\Eqr{efflag15} is sufficient to extract the top quark mass:
\be\label{topmass}
m_t^2=\frac{|M_1^t|^2 s_h^2 c_h^2}{\Pi_{0}^{t}(\Pi_{0}^{q}+\Pi_{1}^{q}\(1+c_{h}^{2}\)+\Pi_{2}^{q}s_{h}^{2}c_{h}^{2})}\approx \xi \frac{|M_1^t|^2 }{\Pi_{0}^{t} \, \Pi_{0}^{q}}.
\ee
Here the form factors are evaluated at $p=0$ which neglects effects of order $m_t^2/m_T^2$ where $m_T$ is the mass of the fermionic resonances in the strong sector. In the last equality we assumed the existence of a hierarchy $\Pi_1^q,\Pi_2^q\ll\Pi_{0}^{q}$. As we will explain below this assumption is guaranteed by partial compositeness.

\subsection{The Higgs potential}
\label{HiggsPotential}

As the couplings of the elementary fermions and gauge bosons break the global symmetry of the strong sector, integrating out their high energy modes at 1-loop or higher order will generate a nonvanishing potential for the Higgs boson. At 1-loop the calculation is the one of  Coleman and Weinberg in Ref.~\cite{Coleman:1973jx}.

The gauge contribution to the Higgs potential is obtained starting from \Eqr{Leffgauge1}. After rotation of the form factors to Euclidean space we obtain
\be\l{CWgauge}
V_g(h)=\dst \f{3}{2}\int\f{d^4 p_{E}}{(2\pi)^4} \left[ 2\log\(1+\f{s_h^2}{4}\f{\Pi_1}{\Pi_0}\)+\log\(1+\f{s_h^2}{4}\f{\Pi_1}{\Pi_0}\f{2\Pi_0+\Pi_0^X}{\Pi_0 +\Pi_0^X}\) \right] \,,
\ee
where $p_{E}^{2}=-p^{2}$ is the Euclidean momentum.  The explicit expression of the form factors in a 5D model is obtained in Section~\ref{Generalities5D}. Similarly, the fermionic contribution to the Higgs potential follows from \Eqr{efflag15}
\be\l{CWfermions}
V_{\rm{top}}\(h\)=\dst  -2 N_{c}\int \f{dp_{E}^{4}}{\(2\pi\)^{4}}\left[\log \left(\Pi^{b_{L}} \right)+\log\left(p_{E}^{2}\Pi^{t_{L}}\Pi^{t_{R}}+\left|\Pi^{t_{L}t_{R}}\right|^{2}\right)\right]\,,
\ee
where again all form factors are functions of the Euclidean momentum $p_{E}$.  The convergence of the integrals in \Eqr{CWgauge} and \Eqr{CWfermions} depends on the details of the model. In the holographic case the form factors turn out to be exponentially decreasing functions of $p_{E}$, while the convergence is only power-like in discretized models. 

To extract the structure of the Higgs potential we use again the relations $\Pi_1\ll \Pi_0$ and $\Pi_1^q,\Pi_2^q\ll\Pi_{0}^{q}$ that allow us to expand the logarithms and write the non-constant part of the potential as
\be\label{higgspot}
V(h)=\alpha c_h^2+\beta s_h^2 c_h^2=(\beta-\alpha)s_h^2-\beta s_h^4 \, .
\ee
The coefficients $\alpha$ and $\beta$ are given by
\bes\l{potentialcoefficients}
\begin{align}
\dst \alpha &= -\f{3}{4}\int\f{d^4 p_{E}}{(2\pi)^4}  \f{\Pi_1}{\Pi_0}\( 1+\f{2\Pi_0+\Pi_0^X}{2\(\Pi_0+\Pi_0^X\)}\)-6N_{c}\int\f{d^{4}p_{E}}{\(2\pi\)^{4}}\f{\Pi_{1}^{q}}{\Pi_{0}^{q}}\,, \vspace{8mm}\\\l{potentialcoefficientsb}
\dst \beta &= \dst -2N_{c}\int\f{d^{4}p_{E}}{\(2\pi\)^{4}}\(\frac{\Pi_{2}^{q}}{\Pi_{0}^{q}}-\f{|M_{1}^{t}|^{2}}{p_{E}^{2} \,\Pi_{0}^{q} \, \Pi_{0}^{t}}\)\,.
\end{align}
\ees
The potential in \Eqr{higgspot} has a minimum for
\be
\f{v^{2}}{f^{2}}\equiv\xi = \sin^{2}\(\f{\<h\>}{f}\)=\f{\beta-\alpha}{2\beta}\,,
\ee
corresponding to a physical Higgs mass
\be
m_{h}^{2}=2\f{\alpha^{2}-\beta^{2}}{\beta f^{2}}=-\frac{8\beta}{f^2}\xi(1-\xi)\,.
\label{mhiggs}
\ee
Assuming the gauge contribution to be subleading we can estimate the expected size of $\alpha$ and $\beta$ using spurionic symmetries and dimensional analysis \cite{Mrazek:2011iu}. The Lagrangian in \Eqr{pctopinvSO5} is formally invariant under the spurious symmetry group $\mathcal G$ described in the Introduction. As the composite sector and the high energy fluctuations of the SM fields are integrated out, the Higgs potential has to satisfy the spurionic symmetry $\mathcal G$ with its $SO(5)$ subgroup being non-linearly realized.
Since no non-trivial invariants can be build out of $\lambda_R P_{R}$, this structure will not enter the potential. For $\lambda_L P_L^\alpha$ there are two possibilities 
\be
I_1 \equiv  (U^\dagger P_L^\alpha P_{L\alpha}^\dagger U)_{55} =\Sigma^T  P_L^\alpha P_{L\alpha}^\dagger\Sigma=1-\frac{3}{4}s_h^2 \,,
\ee
\be
I_2  \equiv  (U^\dagger P_L^\alpha U)_{55} (U^\dagger P_{L\alpha}^{\dagger} U)_{55}= (\Sigma^T P_L^\alpha \Sigma)(\Sigma^T P_{L\alpha}^\dagger \Sigma)=s_h^2c_h^2 \, .
\ee
Both invariants can be generated at 1-loop order and are proportional to $\lambda_L^2$. The leading contribution to the potential will thus be of the form
\be\label{a1a2}
V(h)\approx N_C\frac{m_\psi^4}{16\pi^2}\frac{\lambda_L^2}{g_{\psi}^2}\left(a_1 I_1+a_2 I_2\right) \,,
\ee
where $a_1$, $a_2$ are coefficients of order 1, and $m_\psi\equiv g_{\psi} f$ is the typical mass of the fermionic resonances cutting off the UV divergences in $V(h)$. As $a_1$ is naturally of the same order as $a_2$, a given value of $\xi$ requires a cancellation of order $\xi$ among the parameters. The Higgs mass, on the other hand, is given by
\be\label{higgsmass}
m_h^2\sim N_C\frac{g_{\psi}^2}{2\pi^2}\frac{g_{\psi}^2}{\lambda_R^2} y_t^2 v^2 |a_{2}|(1-\xi) \approx (380~\textrm{GeV})^{2} \frac{1}{\epsilon_R^2}\left(\frac{g_{\psi}}{4}\right)^2 |a_{2}|.
\ee
This implies that, to obtain the measured value of the Higgs mass in a natural way, we need $\epsilon_R\approx 1$ (in which case $t_R$ would be a fully composite state), a small $g_{\psi}$ and $|a_{2}|\sim O(1)$ or a suppressed quartic coupling $|a_{2}|\lesssim 1$. This second possibility will increase the tuning from $\xi$ to $\xi\times|a_2|$ if it is obtained through cancellations between the two pieces in the expression of $\beta$ in Eq.~\eqref{potentialcoefficients}. 
These considerations must be taken with a grain of salt. While the parametric behavior of $m_h$ as a function of the various parameters is a solid prediction, the overall normalization of \Eqr{higgsmass} is only an educated guess. In a specific model, numerical factors of order 1 can conspire to make $|a_2|$ somewhat larger or smaller than 1 in a completely natural way. In the former case the tuning needed to obtain a light Higgs boson would be enhanced while in the latter it would be reduced.
We will show an effect of this kind in our specific example in Section~\ref{Sec:results}.


\section{The 5D construction}\l{Sec:5Dmodel}

\label{Generalities5D}

In order to make detailed predictions from our setup we need a framework where the Higgs potential is calculable. In this paper we follow the holographic approach where finiteness is ensured by locality in an extra dimension. We define our theory in a compact extra dimension $\mathbb R^4\times [0,L]$ with a flat metric
\be
ds^2=\eta_{\mu\nu} dx^\mu dx^\nu-dz^2,
\ee
and with $L^{-1}=O(\textrm{TeV})$.
To compute the effective action for the light degrees of freedom we follow the holographic prescription as described in Ref.~\cite{Panico:2007fq}.\footnote{For a detailed analysis of the holographic techniques and gauge-Higgs unification in a flat extra dimension see Ref.\,\cite{Serone:2009ks}.}

One may question the use of the flat metric in our construction: extra dimensional models require warping in order to properly address the hierarchy problem. It is known however that it is possible to mimic the low energy behavior of a theory in warped space (AdS$_5$ in particular) by a theory formulated in a flat extra dimension with suitable terms added to its action \cite{Scrucca:2003cf,Barbieri:678689, Panico:2010ir}. This is enough to implement features like partial compositeness in flat space. In a warped theory these properties would follow automatically from the dual 4D interpretation.

In Appendix \ref{App5D} we give all the formulas which are necessary to implement our model also on an AdS$_5$ background.

\subsection{Gauge degrees of freedom}\l{App5Dgauge}

The bulk theory obeys a gauged $SO(5)\times U(1)_X$ symmetry\footnote{The gauged symmetry includes color $SU(3)$ which we do not write explicitly.} corresponding to the action
\be\l{gauge5D1}
S_{5D}^g=-\int d^{4}x\int_0^L \frac{dz}{L} \left[\frac{1}{4 g_{5}^{2} } \Tr[F_{MN}^2] + \frac{1}{4g_{X}^{2}}\left(F^{X}_{MN}\right)^2 \right]\, .
\ee
The gauge symmetry is broken to $SU(2)_L\times U(1)_Y$ in the UV ($z=0$) and to $SO(4)\times U(1)_X$ in the IR ($z=L$) by the boundary conditions
 \cite{Barbieri:678689}
\be\label{gaugeBC}
\begin{array}{lll}
F^{a_L}_{\mu5}=F^{a_R}_{\mu5}=F^{X}_{\mu5}=0, &\qquad A^{\hat a}_\mu = 0, &\qquad z=L, \vspace{2mm}\\
A^{a_L}_\mu=W^a_\mu, \qquad A^{3_R}_\mu=A^X_\mu=B_\mu, &\qquad A^{1_R,2_R}_\mu=A^{\hat a}_\mu=0, &\qquad z=0.
\end{array}
\ee
The $z=0$ values of the 5D fields are used as interpolating fields (holographic fields) in the low energy theory.

The bulk gauge symmetry has to be gauge fixed. A useful gauge to adopt is the one in which $A_5$ vanishes along the extra dimension. This is reached by a rotation of the gauge fields by the Wilson line
\be\label{holographicgauge}
\bar g(x,z)= \textrm{P}\left[\exp\left(i\int_0^z dz' A_5^a(x,z') T^a\right)\right] \, ,
\ee
where `P' stands for path ordered exponential and $T^a$ are the $SO(5)$ generators. The issue with $\bar g$ is that it does not reduce to an element of $SO(4)$ on the IR boundary. This problem can be bypassed by enlarging the gauge group at $z=L$ through the explicit introduction of Goldstone bosons. The IR boundary conditions in \Eqr{gaugeBC} are therefore redefined in terms of the rotated field
\be
A_M^{(U)}\equiv U(A_M+i\partial_M) U^\dagger \, ,
\ee
where $U$ is the matrix in \Eqr{GB} and
\be\label{rotation}
\left(A^{(U)}\right)^{\hat a}_\mu(x,L)=0 \, , \qquad \left(F^{(U)}\right)^{\hat a}_{\mu 5}(x,L)=0.
\ee 
This corresponds to an $SO(5)$ gauge transformation by the matrix $U$. Thanks to the transformation properties of $U$ the IR boundary conditions are now invariant under the full $SO(5)$ group and  \Eqr{holographicgauge} can be used to reach the $A_5=0$ gauge. It is important to keep in mind that no new degree of freedom has been introduced in the theory. This is clear since the new scalars in $U$ can be gauge fixed to zero by an appropriate gauge transformation on the IR boundary.
The dependence of the IR action on $U$ can be removed through a bulk gauge transformation $A\to A^{(U^\dagger)}$. As the bulk action is $SO(5)$ invariant, the effect is merely to move the dependence on $U$ to the UV boundary conditions which become
\be\label{holographicgauge1}
A_\mu(x,0)=a_\mu^{(U)}.
\ee
This rotation will affect all the fields in the theory according to their $SO(5)$ representations and determine how the Higgs boson enters the theory.

Localized terms consistent with the reduced gauge symmetry can be added on both boundaries. For the gauge fields we add kinetic terms for the holographic sources
\be\l{gauge5D2}
S_{\text{UV}}^{g} =- \int d^{4}x \left[\frac{1}{4g_2^2} \left(W^a_{\mu\nu}\right)^2 - \frac{1}{4 g_1^2} \left( B_{\mu\nu} \right)^2\right] .
\ee
Following Ref.~\cite{Panico:2007fq} we obtain the effective action for the holographic degrees of freedom. We solve the classical equations of motion for the bulk fields with the boundary conditions specified in \Eqr{gaugeBC} and we get
\be\label{Leffgauge}
\dst S^{g}_{\rm{eff}}  =  \dst S_{\text{UV}}^{g} - \int d^{4}x \left[ \frac{1}{2 g_5^2 L}\Tr[A_\mu\partial_5 A^\mu]+\frac{1}{2 g_X^{2} L}A^X_\mu\partial_5 A^{X\mu}\right] \, .
\ee
The gauge fields can be split into their longitudinal and transverse parts $A_\mu=A^{(l)}_\mu+A^{(t)}_\mu$.
It is easy to show that the solution of the equation of motion for the longitudinal part is
\be
A^{\hat a(l)}(z)=A^{\hat a(l)}(0)\left(1 - \frac{z}{L} \right),\qquad A^{a(l)}(z)=0.
\ee
The $A_5=0$ gauge is reached by rotating the UV boundary values of the gauge fields as in \Eqr{holographicgauge1}. The residual gauge freedom can be fixed by going to the Landau gauge $B_\mu^{(l)}=W^{a\,(l)}_\mu=0$. Now the terms in \Eqr{Leffgauge} corresponding to the longitudinal part of the broken $SO(5)/SO(4)$ gauge fields deliver the kinetic term for the Goldstone bosons 
\be\label{Leffgauge4}
\mathcal{L}_{\text{kin}} =-\frac{1}{2g_5^2 L^2}\Tr[(U^\dagger \partial_\mu U)^2] \,.
\ee
Matching its normalization to \Eqr{sigmamodel} implies the relation
\be\l{Eq:f}
f^2=\frac{2}{L^2}\frac{1}{g_5^2}\,.
\ee
From the remaining part of \Eqr{Leffgauge} one obtains the form factors appearing in \Eqr{gaugeFF}
\be\label{gaugeFF1}
\bry{lll}
\Pi^X_{0}&=&\dst \frac{p^2}{g_1^2}-\frac{p^2}{g_2^2}+\frac{\Pi(p)_V}{g_X^2L},\vspace{2mm}\\
\Pi_0&=&\dst \frac{p^2}{g_2^2}+\frac{\Pi(p)_V}{g_5^2L},\vspace{2mm}\\
\Pi_1&=&\dst 2\frac{\hat \Pi_V(p)-\Pi_V(p)}{g_5^2L}\,,
\ery
\ee
where $\Pi_V(p)=p\tan pL$ and $\hat\Pi_V(p)=-p\cot pL$. Using Eqs.~(\ref{gaugeFF}) and (\ref{gaugeFF1}) we obtain for the gauge couplings
\be\label{gaugecoupling}
\frac{1}{g^2}=\frac{1}{g_2^2}+\frac{1}{g_5^2}\(1-\frac{\xi}{3}\)\approx\frac{1}{g_2^2},\qquad\quad \frac{1}{g^{'2}}=\frac{1}{g_1^2}+\frac{1}{g_X^2}+\frac{1}{g_5^2}\(1-\frac{\xi}{3}\)\approx\frac{1}{g_1^2}\,,
\ee
which also imply $g_{1}\approx g'$, $g_{2}\approx g$ and therefore $g_{1}= g_{2} t_{W}$ with $t_{W}$ the tangent of the weak mixing angle. The first zero of $\Pi_{ab}$ in \Eqr{Leffgauge1} corresponds to the $W$ mass. In the limit $g_2\ll g_5$ this is given by
\be\label{wmass}
m_W^2\approx \frac{1}{L^2}\frac{g_2^2 \xi}{2 g_5^2}.
\ee
The masses of the Kaluza-Klein resonances can be obtained from the zeros of the form factors. The first KK partner of the $W$ has a mass, before EWSB, given by
\be\label{mKK}
M_{KK}=\frac{\pi}{2L}+O\left(\frac{g_2^2}{Lg_5^2}\right).
\ee
Also the $\hat{S}$ parameter is readily computed from \Eqr{Leffgauge1}
\be\label{shat}
\hat S\equiv g^2\Pi_{30}'(0)=\frac{\xi}{3\(1+g_{5}^{2}/g_{2}^{2}\)-\xi}\approx \frac{g_{2}^{2} \xi}{3g_5^2}.
\ee
One can interpret $\hat S$ as originating from the tree level exchange of the massive KK vectors
\be\label{shat2}
\hat S \sim  \frac{m_W^2}{M_{KK}^2}.
\ee
Notice the factor of $\pi^2$ mismatch between Eqs.~\eqref{shat} and \eqref{shat2} when Eq.~\eqref{mKK} is used. This is due to the sum over the whole KK tower ($M_{KK}^{(n)}\sim n\pi/L$) since $\sum 1/n^{2}=\pi^{2}/6$.

Substituting the form factors in Eqs.~(\ref{gaugeFF}) and (\ref{gaugeFF1}) into the potential of \Eqr{CWgauge} and assuming $g_2\ll g_5$, $g_1\ll g_X$ and $g_5\approx g_X$ we can perform the integral explicitly to get
\be\l{gaugepotential}
\bry{lll}
\dst V_{g}(s_{h})
&=&\dst -\f{1}{L^4}\f{63\zeta(3)}{256\pi^2}\(1+\f{t_{W}^{2}}{3}\)\f{g_{2}^{2}}{g_{5}^{2}}c_{h}^{2}.
\ery
\ee
Notice that the gauge contribution to the potential gives a positive mass term for the Higgs and therefore does not break electroweak symmetry \cite{Witten:1983ut,1984NuPhB.234..173V}.

Let us now discuss the role of the UV localized kinetic terms for the gauge fields in a flat extra dimension \cite{Panico:2010ir}. As shown in \Eqr{gaugecoupling}, in the absence of these contributions the kinetic terms of the gauge bosons would come only from the strong sector $1/g^2\approx 1/g_5^2$. In this case we would expect no difference between the electroweak bosons and the other vector resonances of the theory: this is exemplified by the value of $\hat S$ which is of order one without UV localized kinetic terms. Such a scenario is clearly not viable. The inclusion of large kinetic terms on the UV boundary is required to ensure that the SM $W$ and $Z$ are sufficiently weakly coupled to the strong sector. The parameter $g/g_{5}$ can be understood as the degree of mixing between the elementary gauge fields and the composite resonances of the strong dynamics.

When the theory is formulated in AdS$_5$ space the suppression of the mixing between the elementary and the composite sector is guaranteed by the curvature of the metric without the necessity to add localized kinetic terms (see Appendix \ref{App5D}).
The reason for this is the presence of a warping which makes a clear distinction between the UV localized (elementary) and IR localized (composite) states: the small overlap of their wave functions automatically ensures a suppression of their mixing.



\subsection{Fermionic degrees of freedom}\label{sectionfermions}

We introduce two bulk Dirac fermions in $SO(5)\times U(1)_{X}$ representations: \sloppy\mbox{$\Psi_q=\mathbf{14}_{2/3} = \mathbf{1}_{2/3}\oplus\mathbf{4}_{2/3}\oplus\mathbf{9}_{2/3}$} and $\Psi_t=\mathbf{1}_{2/3}$. We work in the massless $b$ quark limit. To properly account for the $b$ quark mass two further multiplets embedding a $q_{L}$ and a $t_{R}$ with $X=-1/3$ have to be introduced with appropriate boundary conditions.
There is some degree of arbitrariness in the definition of the  boundary conditions for the various fermionic components. Our choice is the following\footnote{For the decomposition of the $\mathbf{14}_{2/3}$ representation in terms of SM quantum numbers see Appendix \ref{App5Dfermions}.}
\begin{align}\l{fieldsBC1}
&\dst \Psi_{t}=\(\psi_{tL}(-+)\qquad \psi_{tR}(+-)\)\,,\\
& \Psi_{q}\supset\left\{\bry{ll} \psi^{\(\mathbf{1}\)}_{qL}(--)\qquad\qquad\quad\,\,\, &\dst \psi^{\(\mathbf{1}\)}_{qR}(++)
\\
\psi^{\(\mathbf{4}\)}_{qL}=\(\bry{l} q_{L}'(-+)\\q_{L}(++)\ery\)\quad &\psi^{\(\mathbf{4}\)}_{qR}=\(\bry{l} q_{R}'(+-)\\q_{R}(--)\ery\)\\
\psi^{\(\mathbf{9}\)}_{qL}(-+)\quad &\psi^{\(\mathbf{9}\)}_{qR}(+-)\ery\right.\,
\end{align}
where we take the UV boundary values of the $T^3_R= -1/2$ components of $\psi^{\(\mathbf{4}\)}_{qL}$ and of $\psi_{tR}$ as holographic fields.
Notice that the boundary conditions  on the IR brane commute with the unbroken $SO(4)$ symmetry. The two multiplets are assigned the bulk masses $M_{\Psi_t}$ and $M_{\Psi_q}$ respectively so that the bulk action is given by
\be
S_{5D}^{f}=\int d^{4}x\int_0^L \frac{dz}{L}~ \Tr[\ovl \Psi_q\(i\slashed D+M_{\Psi_q}\)\Psi_q]+ \Psi_t\(i\slashed D+M_{\Psi_t}\)\Psi_t \,,
\ee
where the factor $1/L$ has been introduced to ensure the dimensionality of the 5D fermions to be equivalent to the canonically normalized 4D fields.
At this level the only zero modes are a left-handed $SU(2)_L$ doublet, $q_L$, and a right-handed $SU(2)_L$ singlet, $\psi^{\(\mathbf{1}\)}_{qR}$, both coming from $\Psi_q$.

The IR boundary terms allowed by the unbroken $SO(4)\times U(1)_X$ symmetry at $x=L$ are given by the sum of four different kinetic terms for the four $SO(4)$ representations plus a mass mixing between the two $SO(4)$ singlets:
\be\l{IRaction}
\bry{llll}
\dst S_{\text{IR}}^{f}&=&\dst \int d^{4}x\int_0^Ldz~ \bigg[ &\left( k^t_1\ovl{\psi}_{tL}i\slashed D \psi_{tL}+k^q_1\ovl{\psi}^{\(\mathbf{1}\)}_{qR}i\slashed D \psi^{\(\mathbf{1}\)}_{qR} +~k^{q}_{4}\ovl{\psi}^{\(\mathbf{4}\)}_{qL}i\slashed D \psi^{\(\mathbf{4}\)}_{qL}+k^{q}_{9}\ovl{\psi}^{\(\mathbf{9}\)}_{qL}i\slashed D \psi^{\(\mathbf{9}\)}_{qL}\right) \vspace{2mm} \\ 
&&& \dst + \left( m_{11} \ovl{\psi}^{\(\mathbf{1}\)}_{qR} \psi_{tL}+\textrm{h.c.}\right) \bigg]~\delta(z-L)\,.
\ery
\ee
The parameter $m_{11}$ has the dimension of a mass and controls the mixing between the right-handed component of the singlet inside $\Psi_q$ and the left-handed component of $\Psi_t$. This mixing ensures that the holographic source for  $\psi_{tR}$ has a non vanishing overlap with the right-handed zero mode living in $\Psi_q$. This is crucial because the top Yukawa coupling arises from the kinetic term of $\Psi_q$ after EWSB.\footnote{Notice that we could have exchanged the boundary conditions of $\Psi_t$ with those of $\psi_1^{(\bf{1})}$ with little change in the physics. Another possibility could have been to dispose of the 5D singlet and to use the UV boundary value of $\psi_{qR}^{(\bf 1)}$ as the holographic field. In a model implemented in AdS$_5$ space this possibility requires both the $t_L$ and the $t_R$ to couple to the same operator in the strong sector thus implying $\lambda_L\sim \lambda_R\sim \sqrt{y_t g_\psi}$. This is problematic for two reasons: firstly it does not allow us to reach the limit in which the $t_R$ is fully composite thus disfavoring a light Higgs boson; secondly it creates a tension with EWPT, in particular with the $\hat T$ parameter, which scales like $\hat T\sim\frac{g_\psi^2}{(4\pi)^2}\frac{\lambda_L^4}{g_\psi^4}\xi$. Although there is more freedom in flat space, we discard this possibility to make the comparison with AdS$_5$ more transparent.} The parameters $k_{i}$ are dimensionless and control the magnitude of the IR localized kinetic terms for the four different $SO(4)$ representations. Notice that $k_{1}^{t}$ and the combination $k_{1}^{q}=k_{4}^{q}=k_{9}^{q}$ are $SO(5)$ invariants. To simplify the discussion and without losing any qualitative effect we set $k_{1}^{t}=k_{1}^{q}=0$. Non vanishing values for $k_{4}^{q}$ and $k_{9}^{q}$ are thus needed to break $SO(5)$ on the IR boundary. Moreover a non vanishing difference $k_{4}^{q}-k_{9}^{q}$ is required to break an $SU(9+4)=SU(13)$ accidental global symmetry of the bulk theory under which the fields in the ${\bf 9}$ and those in the ${\bf 4}$ rotate as a single multiplet. In the absence of such a breaking the $\Pi_1^q$ form factor would vanish: the only contribution to $\alpha$ in \Eqr{higgspot} would then come from the gauge sector.

As discussed in the previous section for the gauge sector, realizing partial compositeness in a flat extra dimension requires the introduction of UV localized kinetic terms. We introduce the UV boundary action
\be\label{UVfermions}
S_{\text{UV}}^{f}=\int d^{4}x\int _0^Ldz ~\left[Z_q\,\overline q_Li\slashed D q_L+Z_t\,\overline q_Ri\slashed D q_R\right]\delta(z),
\ee
which contributes to the form factors in \Eqr{fffermions} as
\be
\Delta \Pi_{0}^{q}=Z_q,\quad\Delta\Pi_{0}^{t}=Z_t.
\ee
Assuming $Z_{q}\gg 1$ implies a hierarchy $\tilde{\Pi}_{0}^{q}\gg\Pi_1^q,\Pi_2^q$ which is the one invoked in \Eqr{topmass} and above \Eqr{higgspot}.\footnote{In flat space we will use the notation $\tilde{\Pi}_{0}^{q}=Z_{q}+\Pi_{0}^{q}$, $\tilde{\Pi}_{0}^{t}=Z_{t}+\Pi_{0}^{t}$ as we do in Appendix \ref{App5Dfermions} to distinguish the form factors with and without the localized kinetic terms. Notice that $\tilde{\Pi}_{0}^{q}$ and $\tilde{\Pi}_{0}^{q}$ enter in the Coleman-Weinberg potential of Eq.~\eqref{CWfermions}.} All interactions of the elementary fields with the strong sector get suppressed by a factor $Z_{q,t}^{-1/2}$ which can be interpreted as the mixing $\epsilon_{L,R}\equiv \lambda_{L,R}/g_{\psi}$ between elementary and composite fields (see Introduction). In view of Eq.~\eqref{higgsmass}, to obtain a composite top right ($\epsilon_{R}\approx 1$) we take $Z_t=0$.\footnote{Other UV and IR boundary terms are needed to ensure that the total variation of the 5D action vanishes \cite{Contino:2004cj}. These terms play a crucial role in the calculation of the 4D effective action and are listed in Appendix \ref{App5Dfermions}.}

\subsection{The spectrum of fermionic resonances}
\l{Sec:Spectrum}

In this paragraph we review how the spectrum of fermionic resonances is encoded in the holographic action for the boundary degrees of freedom.
Using the holographic procedure one can obtain the effective Lagrangian in Fourier space for a fermionic bulk field $\Psi$
\be
\overline\Psi(p) \,\slashed p \Pi(p)\, \Psi(-p) \,.
\ee
The function $\Pi(p)$ contains all the information about the masses of the physical states in the theory. If the source $\Psi$ is not dynamical (Dirichlet BC in the UV), holography relates $\slashed p \Pi(p)$ with the two-point function $\langle \ovl{\mathcal O}_\Psi(p)\mathcal O_\Psi(-p)\rangle$ of the operator $\mathcal O_\Psi$ associated to $\Psi$ in the dual 4D picture. This implies that the masses of the physical states are associated with the poles of $\slashed p \Pi(p)$ (the poles in the two point functions of $\mathcal O_\Psi$). If on the other hand the source $\Psi$ is dynamical, it mixes with the bound states interpolated by $\mathcal O_\Psi$ and the masses of the physical states are given by the zeros of $\slashed p \Pi(p)$.
In the following we will have to deal with cases where $\Psi$ is a vector and $\Pi$ a matrix. Here three possibilities occur. If all the sources in $\Psi$ are non dynamical it is sufficient to consider the poles of any of the entries in $\slashed p \Pi(p)$ as all the states with the same quantum numbers in the strong sector are mixed. If all the sources are instead dynamical one has to find the zeros of the various eigenvalues of $\slashed p\Pi( p)$, that is the zeros of their determinant. Finally, in the intermediate situation in which only a subset of sources is dynamical, it is enough to find the zero eigenvalues of the sub-matrix corresponding to these fields.

For the model presented in this paper, the resonance spectrum before EWSB consists of various towers labeled by their $SU(2)_L\times U(1)_Y$ quantum numbers:
\begin{itemize}
\item a ${\bf 1}_{2/3}$ tower with masses given by the zeros of $\slashed p\tilde{\Pi}_{0}^{t}$;
\item a ${\bf 2}_{1/6}$ tower with masses given by the zeros of $\slashed p(\tilde{\Pi}_{0}^{q} + 2 \Pi_{1}^{q})$;
\item a ${\bf 2}_{7/6}$ tower with masses given by the poles of $\slashed p(\Pi_{0}^{q} + 2 \Pi_{1}^{q})$;
\item a ${\bf 9}_{2/3}$ tower under $SO(4)$, decomposing as $\mathbf 3_{5/3}\oplus \mathbf 3_{2/3}\oplus \mathbf 3_{-1/3}$ under $SU(2)_L\times U(1)_Y$ with masses given by the poles of $\slashed p\Pi_{0}^{q}$.
\end{itemize}
After EWSB the physical states are organized according to their electric charge:
\begin{itemize}
\item a charge 2/3 tower with masses given by the zeroes of 
\be\nonumber
p^{2} \tilde{\Pi}_{0}^{t} \(\tilde{\Pi}_{0}^{q} + \Pi_{1}^{q} \(1 + c_{h}^{2} \) + \Pi_{2}^{q} c_{h}^{2} s_{h}^{2} \) - |M_{1}^{t}|^{2} s_{h}^{2} c_{h}^{2} ,
\ee
where the first zero corresponds to the top quark mass;
\item a charge -1/3 tower with masses given by the zeros of $\slashed p(\tilde{\Pi}_{0}^{q} + 2 \Pi_{1}^{q} c_{h}^{2})$, where the massless pole corresponds to the bottom quark;
\item a charge 5/3 tower with masses given by the poles of $\slashed p(\Pi_{0}^{q} + 2 \Pi_{1}^{q} c_{h}^{2})$;
\item charge 8/3 and -4/3 towers with masses given by the poles of $\slashed p\Pi_{0}^{q}$.
\end{itemize}
\medskip
All the form factors appearing in these expressions are given in Appendix \ref{App5Dfermions}.

Before closing this section it is important to discuss a crucial feature of holographic composite Higgs models, which pertains the relation between the two couplings $g_\rho$ and $g_{\psi}$ described in the Introduction. They can be defined schematically by their relation to the masses of bosonic and fermionic resonances of the theory: $m_\rho\equiv g_\rho f$ and $m_\psi\equiv g _\psi f$. While these two couplings can be split in generic 4D constructions \cite{Panico:1359049, Redi:qi,Redi:2012vq}, in a 5D theory the Kaluza-Klein masses are all set by a single mass scale, $L^{-1}$.  For the coupling among vectors, using \Eqr{mKK} for the mass of the first KK state and \Eqr{Eq:f}, one gets
\be
g_\rho\equiv \frac{\pi}{2\sqrt 2}g_5. 
\ee
The fermionic spectrum and thus the coupling $g_\psi$ are more model dependent. Considering a 5D bulk fermion with $M_\Psi=0$ whose right-handed component has $(+,-)$ BC we find again
\be
g_\psi\equiv \frac{\pi}{2\sqrt 2}g_5.
\ee
Thus in general
\be
g_{\rho}= g_{\psi}= g_{5}\,.
\ee
This has an important consequence in view of \Eqr{higgsmass} since $g_{\psi}$ is now identified with $g_\rho$. Barring accidental cancellations in the quartic coupling of the Higgs and assuming the most favorable situation of a composite $t_R$ ($\epsilon_R\sim 1$), $g_\rho$ has to be small to accommodate the observed value of $m_h$. This generates a tension with the value of the $\hat S$ parameter (see \Eqr{shat})
\be
\hat S\approx \xi\frac{g^2}{3 g_\rho^2}\approx 10^{-3}\left(\frac{\xi}{0.1}\right)\left(\frac{4}{g_\rho}\right)^2.
\label{Svectors}
\ee
We will come back to this issue in the next Section.


\section{Numerical analysis}\l{Sec:results}


In this section we present a numerical analysis of the model focusing on the Higgs potential and on the fermionic spectrum.
We begin by discussing the parametric behavior of the various observables we are interested in and later present our numerical procedure.

As we are mainly interested in the composite $t_R$ case we first describe the ideal situation in which $Z_q\gg 1$ and $\tilde{\Pi}_{0}^{q}\approx Z_q$. The parameters of the 5D model can be related using the expressions for the weak coupling $g$, the $W$, top and Higgs masses. 
The Higgs potential has the following parametric behavior
\be\label{potentialpar}
V(h)=\frac{1}{L^4}\frac{N_C}{(4\pi)^2Z_q}V_h(\{p_i\};\xi)+V_g(h) \, ,
\ee
where $V_g(h)$ is the gauge contribution given in \Eqr{gaugepotential}. $p_i$ is the set of dimensionless parameters $M_q L,\,M_t L,\, m_{11}L,\, k_1^t,\,k_1^q,\,k_4^q,\,k_9^q$. If we can neglect $V_g$ (that is if $g_{5}$ is sufficiently large) the value of $\xi$ can be determined as soon as the $p_i$ are fixed.
In the limit $g_{5}\gg g$\footnote{These various relations hold with good accuracy as long as $g_{5}\gtrsim 1.5$.} the $W$ mass is given by \Eqr{wmass}.
The top mass can be read off \Eqr{topmass} and, in the large $Z_q$ limit, can be written as
\be\l{topmass2}
m_{t}^{2}=\f{1}{L^{2}}\frac{\xi}{Z_{q}} F_t(\{p_i\}) \, ,
\ee
where $F_t$ is a dimensionless function of the various $p_i$.
Our final input, the Higgs mass, follows from \Eqr{potentialpar}
\be\label{mhscaling}
m_h^2=\frac{1}{L^2}\frac{N_Cg_{5}^{2}}{(4\pi)^2}\frac{\xi}{Z_q}F_h(\{p_i\}).
\ee
For a given set of $p_i$ it is then possible to fix $g_{5}$, $Z_q$ and $L$ to reproduce the three input values $m_W$, $m_t$ and $m_h$.

In order to include the effect of a non vanishing gauge potential and to allow for smaller values of $Z_q$ we scan randomly and uniformly over the parameters $p_i$ and over both $g_{5}$ and $Z_q$. For each point we evaluate $\xi$ and compute the ratios $m_t/m_W$ and $m_h/m_W$. We discard points which do not fall into the interval defined by  
\be
\bry{ccc} \label{passscan}
m_W & = & 81\,{\textrm{GeV}},\\
148\,{\textrm{GeV}}\leq& m_t & \leq 154\,{\textrm{GeV}} \footnotemark  ,\\
120\,{\textrm{GeV}}\leq& m_h & \leq 130\,{\textrm{GeV}}.
\ery
\ee
\footnotetext{The central value corresponds to the top quark mass renormalized at the scale $\mu=1\,{\textrm{TeV}}$, \sloppy\mbox{$m_t=150.7\pm 3.4\,{\textrm{GeV}}$}.}
We choose the following range for the parameters $p_{i}$
\be\label{scanflat}
\bry{ccl}
M_q L,\, M_t L &:& (-2 \div 2),\vspace{0.5mm}\\ 
m_{11}L&:&(0.3 \div  2),\vspace{0.5mm}\\
k_4^q,\, k_9^q&:&(0 \div 2),\vspace{0.5mm}\\
\sqrt{Z_q}&:&(0 \div 10),\vspace{0.5mm}\\
g_5&:&(1 \div 9).
\ery
\ee
To simplify the scan we fix $k_{1}^{t},\, k_{1}^{q}=0$ since these parameters are not expected to modify the qualitative picture as we explained in Section \ref{sectionfermions}. We furthermore restrict $m_{11}$ to positive values as it only enters through the combination $|m_{11}|^{2}$. 

\begin{figure}[t]
\begin{center}
\includegraphics[scale=0.54]{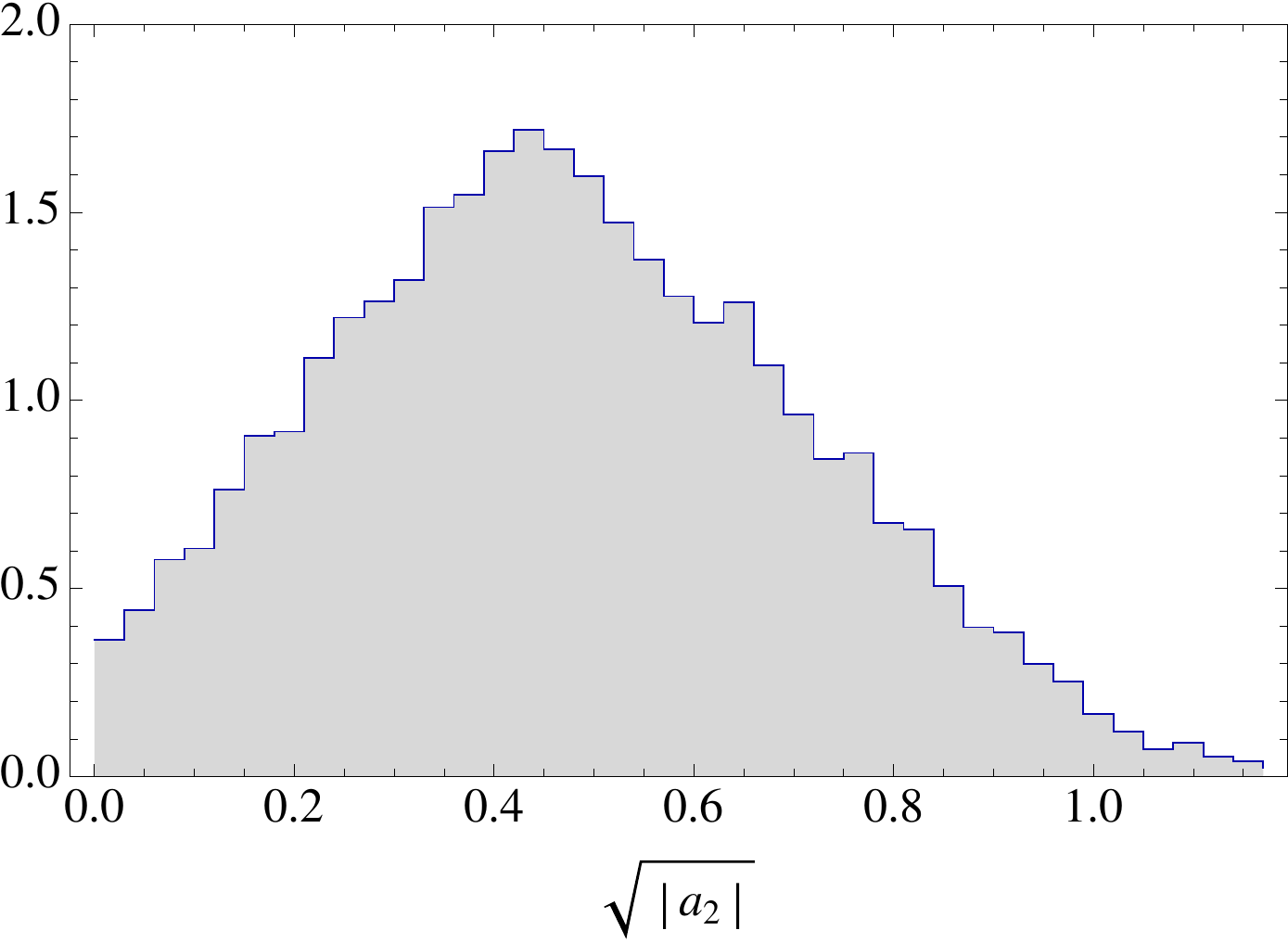}
\end{center}
\vspace{-5mm}
\caption{ \small
Distribution of the values of $\sqrt{|a_2|}$ as defined in \Eqr{higgsmass}. The parameter range is the one defined in \Eqr{scanflat}. We only keep points with negative $a_2$.
}\label{fig:a2prior}
\end{figure}

In order to conform to the general discussion at the end of Section \ref{HiggsPotential} it is necessary to understand the typical size of the coefficient $a_2$ defined in \Eqr{a1a2}. To do so we scan over the range defined by \Eqr{scanflat} and calculate $a_2$ without imposing any additional requirement. This will give the correct normalization of \Eqr{a1a2} within our model. The results are shown in Fig.~(\ref{fig:a2prior}). We selected only points in which $a_2$ is negative (the distribution is basically symmetric around 0). The distribution is quite broad but clearly peaks around  $\sqrt{|a_2|}\sim 0.4$.\footnote{$|a_1|$ has a similar distribution. We have also checked the suppression in the potential does not correlate with the presence of anomalously light fermionic resonances in the spectrum \cite{Matsedonskyi:2012ws}. This could happen if this suppression  was due to the presence of light top partners cutting of the quadratic divergences of the Higgs potential at a lower scale.} Thus, it is more appropriate to rewrite the NDA estimate of the Higgs mass in \Eqr{higgsmass} by factoring out the normalization of $a_2$
\be\label{higgsmassprior}
m_h^2\approx (150~\textrm{GeV})^{2} \frac{1}{\epsilon_R^2}\left(\frac{g_{\psi}}{4}\right)^2 |a_{2}|.
\ee
If the constraints that we impose on the parameter space to obtain the correct values for $\xi$ and the top mass do not push $a_2$ to the right tail of its distribution, it is clear that the tension of the model with the $\hat S$ parameter will be substantially relieved compared to the naive expectation.

\subsection{Results}

\begin{figure}[t]
\begin{center}
\hspace{-4mm}\includegraphics[scale=0.47]{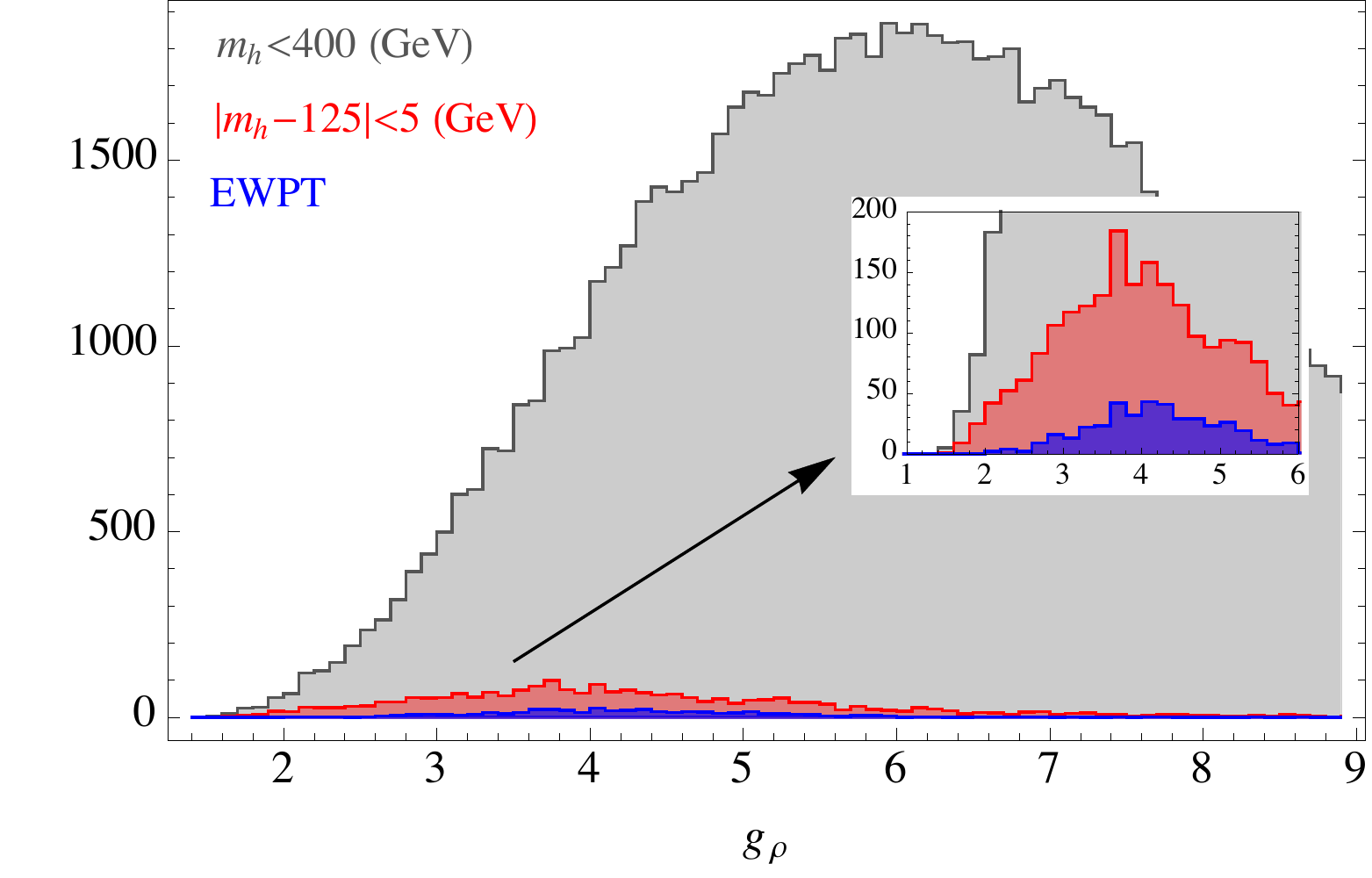}~~~~~~\includegraphics[scale=0.47]{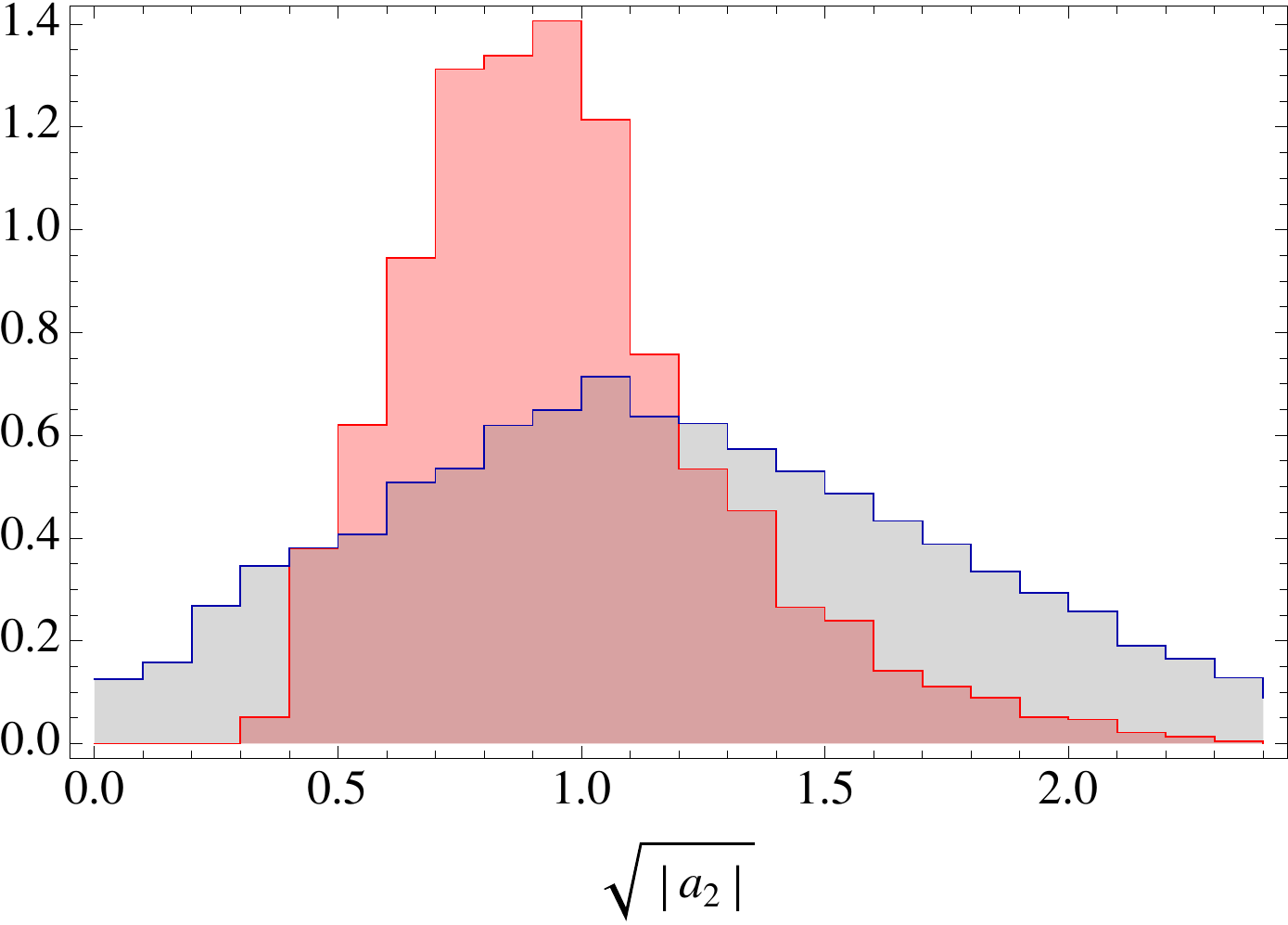}
\end{center}
\vspace{-5mm}
\caption{ \small
Left panel: distribution of points coming from the scan with parameter ranges given by \Eqr{scanflat} as a function of $g_\rho$. Gray: points with $115\,\textrm{GeV}<m_h<400\,\textrm{GeV}$; red:  points with $120\,\textrm{GeV}<m_h<130\,\textrm{GeV}$; blue: points with $120\,\textrm{GeV}<m_h<130\,\textrm{GeV}$ which are consistent with the electroweak fit at 99\% C.L. assuming a $10^{-3}$ positive contribution to $\hat T$. Right panel: distribution of the values of $\sqrt{a_2}$ as defined in \Eqr{a1a2} and \Eqr{higgsmassprior}.
}\label{fig:distgrho}
\end{figure}

In the left panel of Fig.~(\ref{fig:distgrho}) we show the distribution of points passing the various cuts in \Eqr{passscan} as a function of $g_\rho$. The gray histogram includes all points for which \mbox{$m_h<400$ GeV}, while the red distribution requires $m_h$ to fall inside the interval \sloppy\mbox{$120\div130$ GeV}. As expected from the scaling of $m_h$ (see Eqs.~(\ref{higgsmass}) and(\ref{higgsmassprior})), $g_\rho$ cannot be too large and its distribution is peaked between 3 and 4. In the right panel of Fig.~(\ref{fig:distgrho}) we show the distribution of the parameter $\sqrt{|a_2|}$ as defined in \Eqr{higgsmassprior} and compare it with the one already shown in Fig.~(\ref{fig:a2prior}), taking into account the new normalization of $|a_2|$. The scan thus selects values with $a_2\sim 1$, that is in its natural range.

We estimate the tuning of the model as follows. Among all points in which the experimental inputs are reproduced, we pick those satisfying the experimental constraint from the EWPT. These points represent a fraction of the total, and this fraction is what we denote as the tuning of the model. This definition measures the size of the region in parameter space (according to the measure defined by \Eqr{scanflat}) which is left after the various experimental constraints are imposed.
In this way we automatically take into account that unnatural conspiracies among parameters, which are needed to satisfy the experimental constraints, occur rarely \cite{Strumia:2011by}.

Our measure of the tuning differs from the one in Ref.~\cite{Panico:2012vr}  where the more common definition in terms of the logarithmic derivative of $m_Z^2$ with respect to the various parameters was used \cite{Barbieri:1988p1552}.\footnote{The relation between the two criteria is particularly transparent in Supersymmetry \cite{Giudice:961767}. Here one can write a simple relation between the soft parameters at the scale at which they are generated and $m_Z$:
\be
m_Z^2=a M^2+b m^2-c|\mu|^2.\notag
\ee
$M$ and $m$ are the common gaugino and scalar masses at the high scale and $a,b,c>0$. The log-derivative tuning associated to $M$ is given by $\Delta=a M^2/m_Z^2$. Notice that in the $(x\equiv\tfrac{M^2}{|\mu|^2},y\equiv\tfrac{m^2}{|\mu|^2})$ plane this can be interpreted by saying that for a fixed lower bound on $\Delta_{\rm exp}$ (implied for instance by an experimental lower bound $M_{\rm exp}$ on the gluino mass), the only allowed region of the $(x,y)$ plane is the one bounded by the two lines $a x+by-c=0$ (on which $m_Z$ vanishes) and $a x+by-c=a x/{\Delta_{\rm exp}}$. For $a\gtrsim b$ this slice covers a fraction of order $b/(a\Delta_{\rm exp})$ of the whole plane. If we randomly scan the $(x,y)$ plane fixing $\mu$ to reproduce the $Z$ mass, a fraction $b/(a\Delta_{\rm exp})$ of points will satisfy the experimental bound $M>M_{\rm exp}$. This relates the criterion we adopt to the one using the local logarithmic derivative estimate.} 

Requiring the points satisfying \Eqr{passscan} to be consistent with the LEP electroweak fit at 99\% C.L., we remain with the fractions shown in Table \ref{fracpoints}. In this subset, the value of $\xi$ is always below 0.04, corresponding to $f>1.2$\,TeV. Notice that the fraction of points remaining ($\sim 5\%$) is consistent with a measure of the tuning given just by $\xi$. This is the case because as discussed above we do not need any additional suppression in the Higgs quartic coupling in order to get a light Higgs mass and to cope with the constraints on the $\hat S$ parameter. 

Before moving on we notice that by artificially adding an additional positive contribution to $\hat T=10^{-3}$ the fraction of allowed points grows from $5\%$ to roughly 18\% (see the blue histogram in the left panel of Fig.~(\ref{fig:distgrho})). This contribution to $\hat T$, corresponding to one tenth of the top contribution in the SM, is an optimistic estimate of the corrections coming from loops of heavy top partners.\footnote{These contributions are finite and computable in our model. See Ref.~\cite{Agashe:2005p372} for a discussion.}

\begin{figure}[t]
\begin{center}
\includegraphics[scale=0.7]{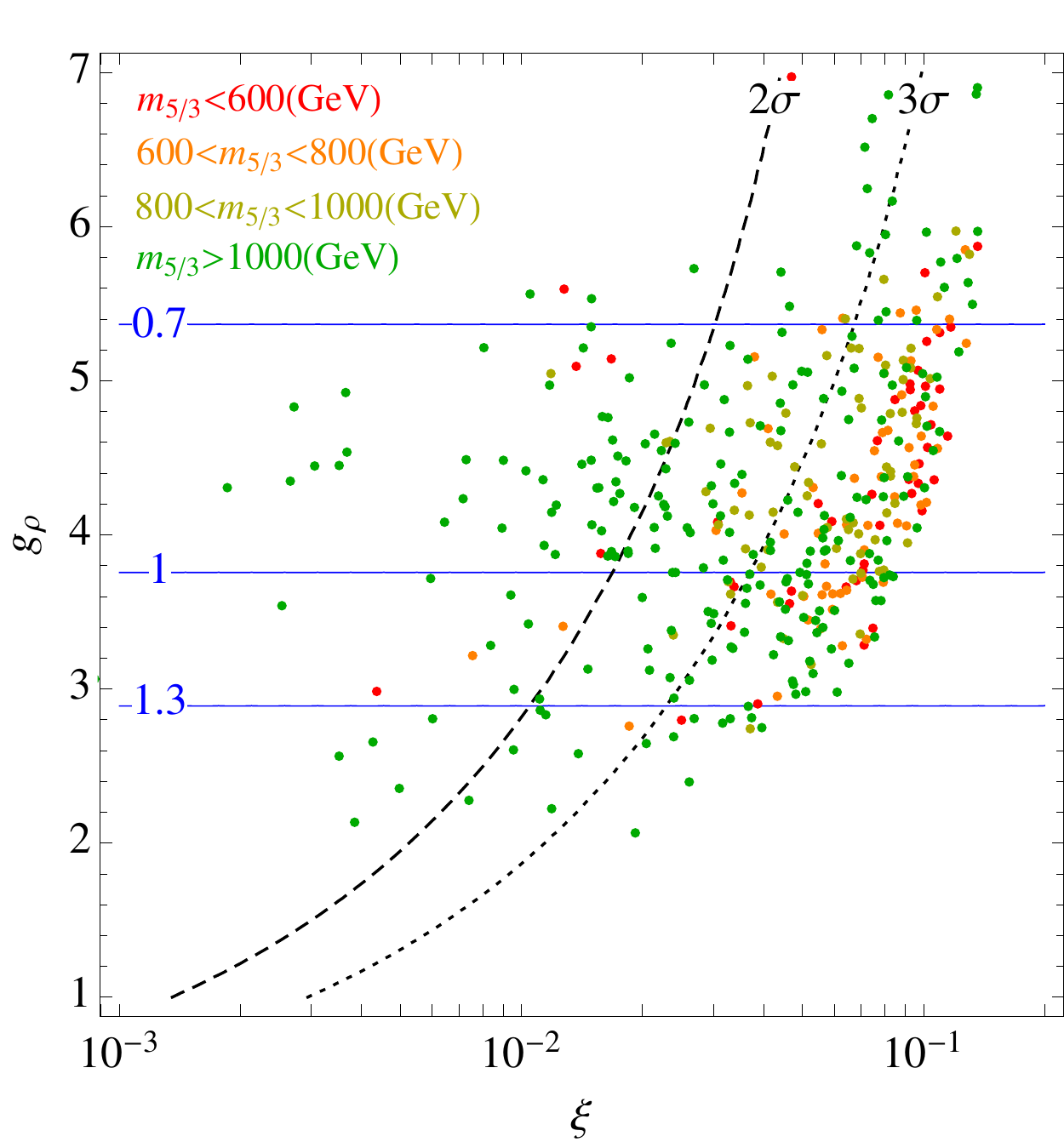}
\end{center}
\vspace{-5mm}
\caption{ \small
Scatter plot in the $(\xi, g_\rho)$ for those points that can be made consistent with the EWPT at 99\% C.L. assuming a $\Delta\hat T=10^{-3}$. The black lines are the $2\sigma$ and $3\sigma$ contours from the LEP EW fit. Blue lines indicate the suppression in $\sqrt{a_2}$ which is necessary to achieve $m_h=125$\,GeV according to the NDA estimate of the Higgs boson mass of \Eqr{higgsmassprior}. The color of the points indicates the mass of  lightest charge 5/3 top partner in the spectrum (see figure).
}\label{fig:points}\vspace{4mm}
\end{figure}
\begin{table}[t]
\begin{center}
\begin{tabular}{c|cc}
 &{\small EWPT  }& {\small{EWPT (+$\Delta\hat T=10^{-3}$)}}\\
 \hline
 \% & $4.5\pm 0.4$& $18\pm 1$
 \end{tabular}
 \end{center}
 \vspace{-5mm}
 \caption{\small Fraction of points where the experimental inputs in \Eqr{passscan} are reproduced which are allowed at 99\% C.L. by the EW fit as described in the text.}\label{fracpoints}
 \end{table}

Figure \ref{fig:points} shows the distribution of points in the ($\xi$, $g_\rho$) plane. We included all points that can be made consistent with EWPT at the 99\% C.L. with the assumption of an additional contribution $\Delta\hat T=10^{-3}$. Red and orange points have a spectrum which contains a charge 5/3 top partner which is lighter than 800\,GeV. Any detailed study of the experimental status of these points is beyond the scope of this work, nevertheless we believe that they should be severely constrained by the available experimental searches \cite{ATLAS-CONF-2012-130, CMS-PAS-SUS-12-027}. The blue contours show the necessary suppression in $\sqrt{a_2}$ according to the NDA estimate of the Higgs boson mass. Fig.~(\ref{fig:points}) is puzzling. If we believe the NDA formula $m_\psi\sim g_\rho f$ we should expect a much heavier spectrum than what we find in Fig.~\ref{fig:points}, as $g_\rho=4$ and $\xi=0.1$ give $m_\psi=3$\,TeV!

\begin{figure}[t]
\begin{center}
\hspace{-1mm}\includegraphics[scale=0.5]{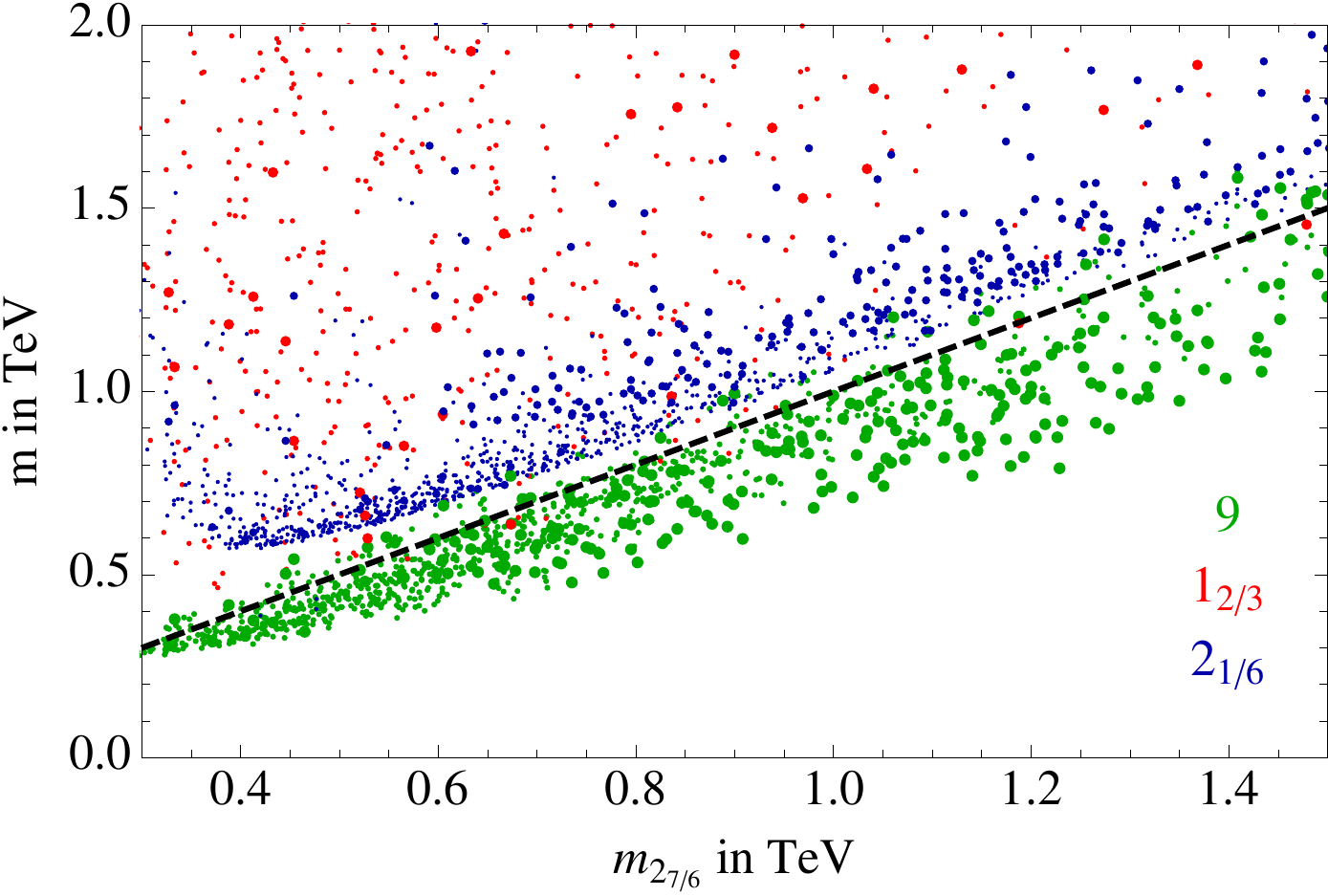}~~~~~~~~~\includegraphics[scale=0.48]{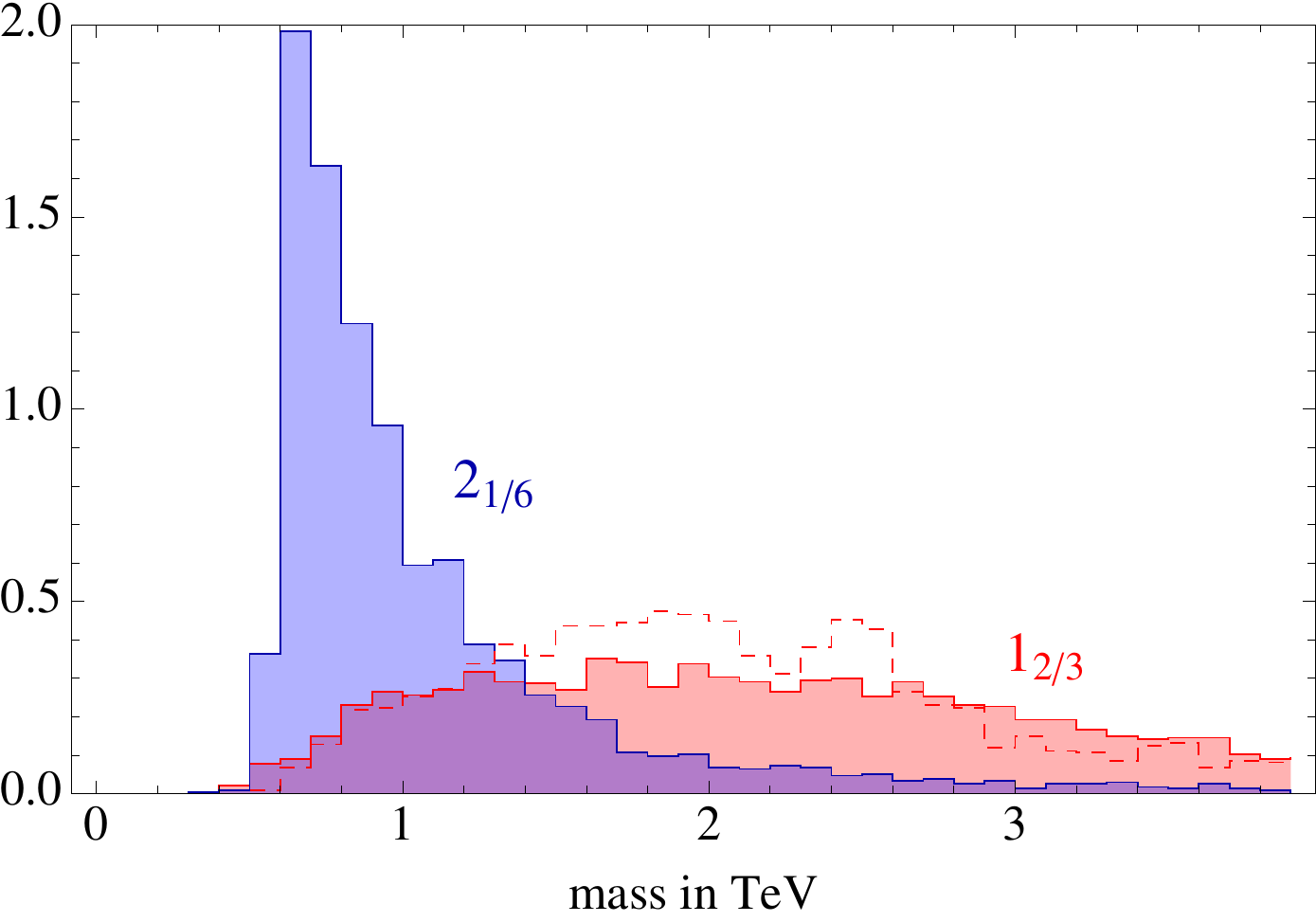}
\end{center}
\vspace{-5mm}
\caption{ \small
Left panel: Distribution of fermionic resonance masses before EWSB according to their quantum numbers: $\mathbf 9$ in green, $\mathbf{2}_{1/6}$ in blue and ${\bf 1}_{2/3}$ in red. Thicker points are those passing the EWPT constraint. Right panel: mass distributions for the ${\bf 2}_{1/6}$ (blue) and ${\bf 1}_{2/3}$ (red) $SO(4)$ representations without considering EWPT. The red dashed histograms is the expected mass distribution from the NDA scaling $m_\psi=g_\rho f$.
}\label{fig:distmasses}
\end{figure}

The solution of this puzzle is shown in Fig.~(\ref{fig:distmasses}). In the left panel we show the fermionic spectrum before EWSB of all points that reproduce the experimental inputs (thicker dots are for points also satisfying EWPT). A definite hierarchy $m_{\mathbf 9}< m_{\mathbf{2}_{7/6}}<m_{\mathbf{2}_{1/6}}\ll m_{\mathbf{1}_{2/3}}$ emerges from the figure. In the right panel of the same figure we compare the mass distribution of the $SO(4)$ representations $\mathbf{1}_{2/3}$ (shaded red) and $\mathbf{2}_{1/6}$ (shaded blue) before requiring EWPT. The red-dashed histogram shows the distribution of masses obtained from the NDA estimate $m_\psi\sim g_\psi f$. It is clear that, while the singlet matches reasonably well with the expectations, the other three $SO(4)$ representations are much lighter.

To understand this behavior it is necessary to inspect how the top mass is obtained in terms of the model parameters. Using the approximation in \Eqr{topmass} and the form factors in Eqs.~\eqref{FFAdS}, one finds that for large positive $M_{\Psi_q}$ the top mass is exponentially suppressed, while for $M_{\Psi_q}<0$
\be\l{mtconstraint}
\frac{m_t^2}{m_W^2}=\frac{g_5^2}{g_2^2}\frac{5|M_{\Psi_q}L|}{Z_q+e^{2|M_{\Psi_q} L|}k_9^q}.
\ee
Reproducing the correct $m_t/m_W$ ratio thus fixes $M_{\Psi_q}\lesssim -1$.
Next, in the limit of large $Z_q$ and vanishing $Z_t, k_4^q$, $k_9^q$, the KK masses of the $\mathbf 9$, ${\mathbf{2}}_{7/6}$ and ${\mathbf{2}}_{1/6}$ towers are all given by the zeros of
$M_{\Psi_q}+\omega_q \cot \omega_q L$ while those of the ${\mathbf{1}}_{2/3}$ tower are given by the zeros of $M_{\Psi_t}+\omega_t \cot \omega_t L$. The approximate expression for the first KK mode mass can be written as
\be\l{approximateKK1}
\bry{lllll}
\dst p\sim 2 |M_{\Psi_{i}}|e^{-|M_{\Psi_{i}}|L}\,, &\qquad&  M_{\Psi_{i}}L\lesssim 0\,,\vspace{3mm}\\
\dst \sqrt{\f{\pi^{2}}{4L^{2}}+M_{\Psi_{i}}^{2}}< p < \sqrt{\f{\pi^{2}}{L^{2}}+M_{\Psi_{i}}^{2}} \,, &\qquad& M_{\Psi_{i}}L\gtrsim 0\,.
\ery
\ee
We see that since $M_{\Psi_{q}}$ should be large and negative to reproduce the correct top mass (see Eq.~\eqref{mtconstraint}) we expect the masses of the first resonances in the $\mathbf 9$, ${\mathbf{2}}_{7/6}$, ${\mathbf{2}}_{1/6}$ towers  to be parametrically (exponentially) suppressed with respect to the mass of the first KK mode in the ${\mathbf{1}}_{2/3}$ tower, whose corresponding bulk mass $M_{\Psi_{t}}$ is unconstrained.

We now explain the smaller splittings giving $m_{\mathbf 9}< m_{\mathbf{2}_{7/6}}<m_{\mathbf{2}_{1/6}}$. First of all, the splitting between the ${\bf 9}$ and the ${\bf 2}_{7/6}$ is proportional to the difference $(k_4^q-k_9^q)$. Expanding the relevant form factors around the point $k_4^q-k_9^q=0$ explicitly shows that $m_{\mathbf 9}-m_{\mathbf 2_{7/6}}$ is positive when $k_4^q-k_9^q$ is positive. Next we rewrite the Higgs potential in \Eqr{higgspot} separating the fermionic and gauge contributions to $\alpha$ and $\beta$:
\be\label{higgspot1}
V(h)=(-\alpha_f-\alpha_g+\beta_f)s_h^2-\beta_f s_h^4.
\ee
The coefficient $\alpha_g$ is negative, $\beta_f$ is always negative in the region we consider, while $\alpha_f$ can have both signs depending on $k_4-k_9$. A viable electroweak symmetry breaking requires the coefficient of the $s_h^2$ term to be negative. The Higgs mass input needs a small $g_5$ which enhances the electroweak symmetry preserving gauge contribution. As the size of $\beta$ is basically fixed by the top and Higgs mass measurements, EWSB typically requires an extra positive contribution to $\alpha$ (see the minus sign in \Eqr{higgspot1}), favoring $k_4< k_9$ and thus $m_{\mathbf 9}<m_{\mathbf 2_{7/6}}$.

\begin{figure}[t]
\begin{center}
\hspace{-4mm}\includegraphics[scale=0.47]{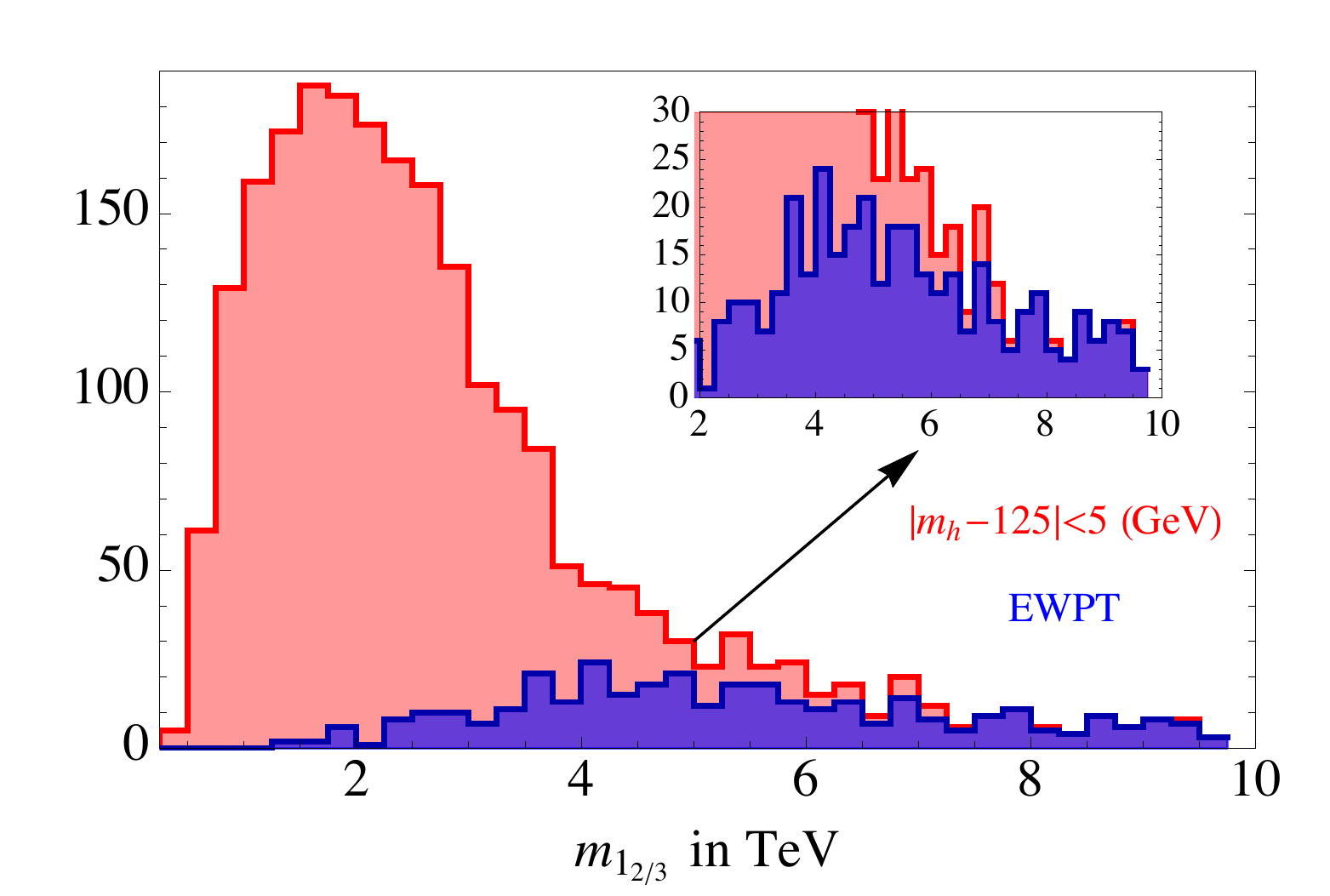}~~~~~~\includegraphics[scale=0.47]{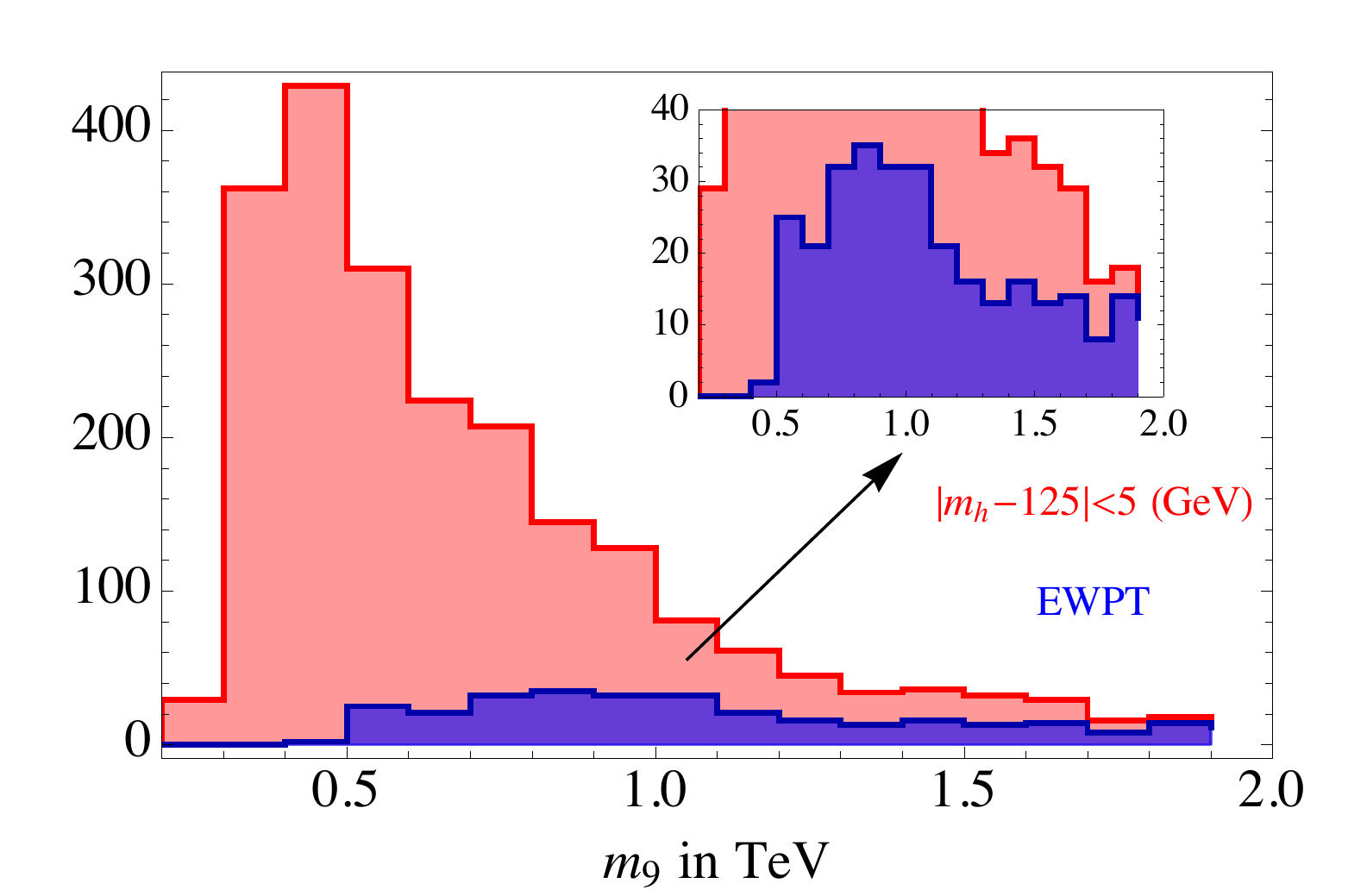}\\
\hspace{-4mm}\includegraphics[scale=0.47]{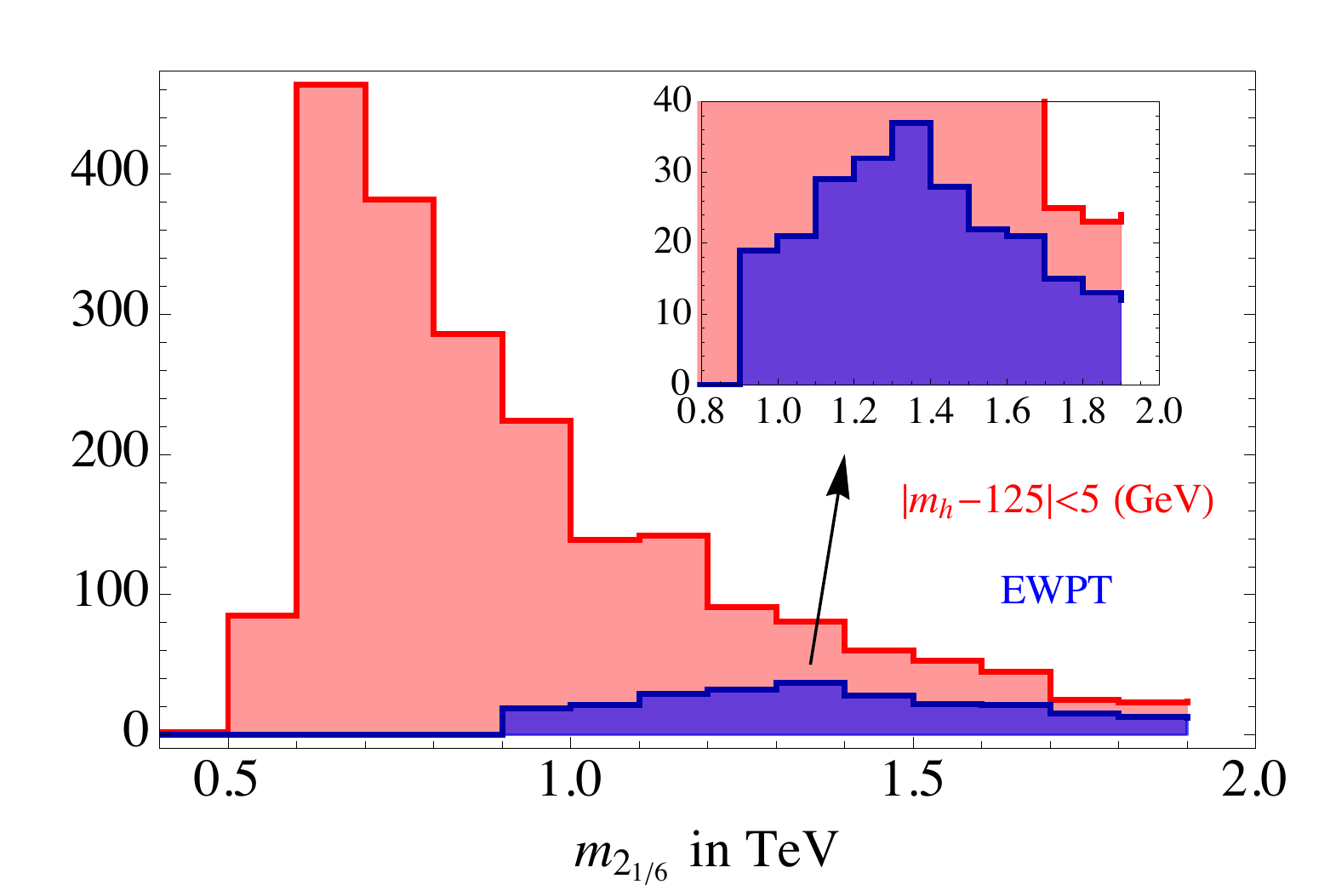}~~~~~~\includegraphics[scale=0.47]{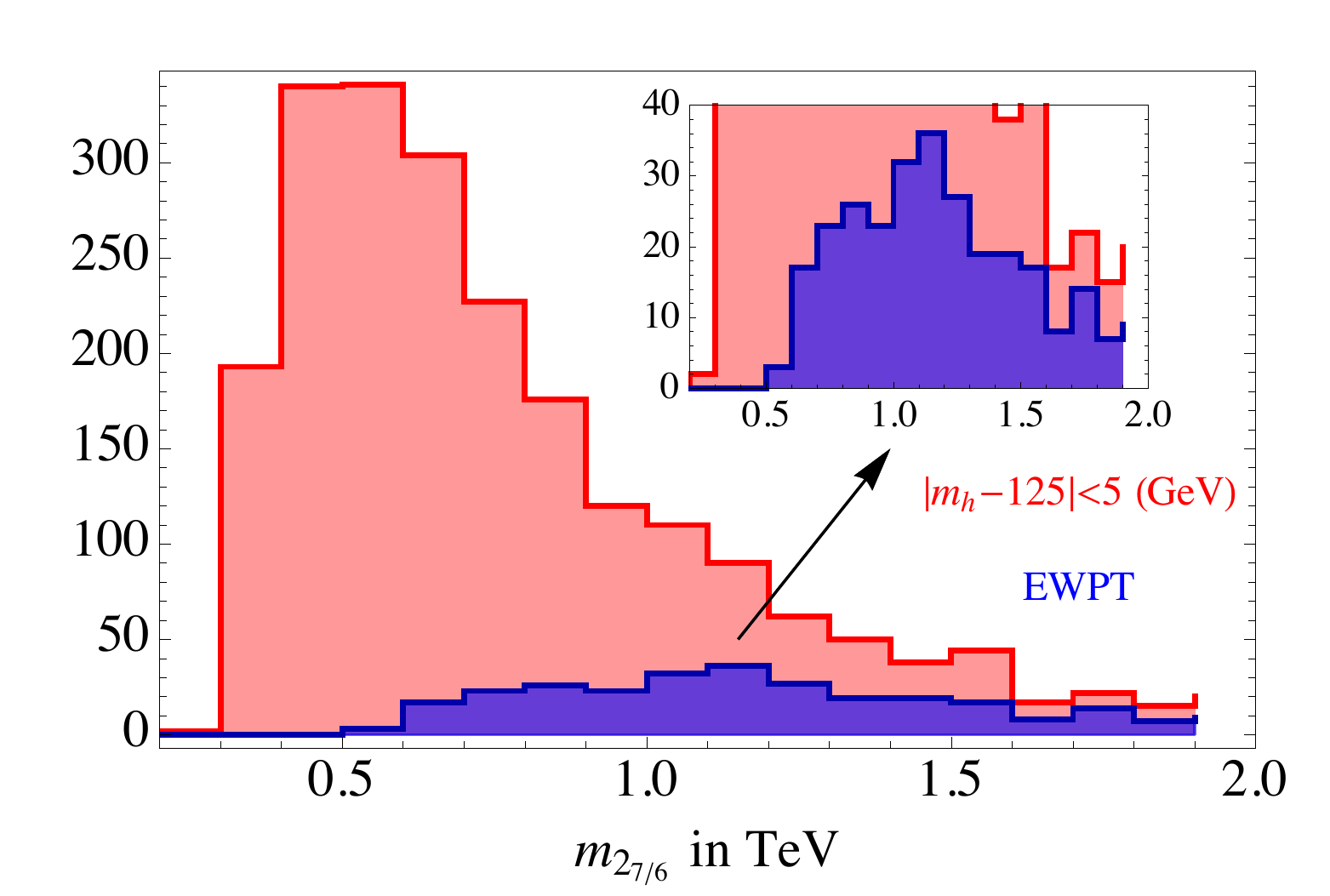}
\end{center}
\vspace{-5mm}
\caption{ \small
Mass distributions of the ${\bf 1}_{2/3}$ (top left), ${\bf 9}$ (top right), ${\bf 2}_{1/6}$ (bottom left) and ${\bf 2}_{7/6}$ (bottom right) obtained from the scan with parameter ranges given by \Eqr{scanflat}. Red: points with $120\,\textrm{GeV}<m_h<130\,\textrm{GeV}$; blue: points with $120\,\textrm{GeV}<m_h<130\,\textrm{GeV}$ which are consistent with the electroweak fit at 99\% C.L. assuming a $10^{-3}$ positive contribution to $\hat T$. 
}\label{fig:masses}
\end{figure}

The mass difference between the ${\mathbf 2}_{7/6}$ and the ${\mathbf 2}_{1/6}$ towers is due to the different UV boundary conditions for these fields. The zeros of $\tilde{\Pi}_{0}^{q}$ give the masses of the ${\mathbf 2}_{1/6}$ while the poles of $\Pi_{0}^{q}$ give the masses of the ${\mathbf 2}_{7/6}$. It is clear that in the large $Z_q$ limit the masses will coincide. In particular, from the explicit form of $\tilde{\Pi}_{0}^{q}$ for $k_9^q=0$ (see Appendix \ref{App5Dfermions}),
\be\label{psiq}
\tilde{\Pi}_{0}^{q}=Z_q+\frac{1}{M_{\Psi_q}L+\omega_qL \cot \omega_qL} \, ,
\ee
it follows that the states in the ${\bf 2}_{7/6}$ are always heavier than those in the ${\bf 2}_{1/6}$. A summary of the mass distributions of the fermionic resonances before (red) and after (blue) requiring the electroweak fit at 99$\%$ C.L. (with an additional positive contribution $\Delta T=10^{-3}$) is given in Figure \ref{fig:masses}.  We show the distributions for the different $SO(4)$ representations ${\bf 1}_{2/3}$ (top left), ${\bf 9}$ (top right), ${\bf 2}_{1/6}$ (bottom left) and ${\bf 2}_{7/6}$ (bottom right).

\section{Outlook and conclusions}\l{Sec:conclusion}

The main challenge of composite Higgs models to naturally describe electroweak symmetry breaking are the constraints coming from precision electroweak physics. The tension of these theories with electroweak precision tests is the one of old fashioned Technicolor models and is due to the unavoidable existence of heavy spin-1 states mixing with the SM gauge bosons. This mixing corrects the $\hat S$ parameter by an amount
\be\label{outlook:S}
\Delta \hat S\sim \frac{m_W^2}{m_\rho^2} \, ,
\ee
requiring $m_\rho\gtrsim 2.6$\, TeV. 

The masses of the spin-1 resonances $m_\rho\sim g_\rho f$ are controlled by two parameters, their coupling $g_\rho$ and the decay constant $f$ determining the non-linear interactions of the composite Higgs. As we described in the text, pushing $f$ to large values implies an upper bound on the tuning of the model of order $\xi\equiv v^2/f^2$. Before Higgs discovery, electroweak physics was pointing towards a strongly coupled theory, $g_\rho\sim 4\pi$, and $\xi\lesssim 0.1$ was needed to regulate the infrared contributions of the composite Higgs to electroweak observables.

The measurement of the Higgs boson mass, $m_h\approx 125$\,GeV, adds extra information to this picture. The generic expectation of the Higgs boson mass in composite Higgs models can be given in terms of the mass of the fermionic resonances which regulate the UV divergence of the Higgs potential:
\be\label{outlook:higgsmass}
m_h\sim 125\,{\textrm{GeV}}\left(\frac{m_\psi}{1.0\,{\textrm{TeV}}}\right)\sqrt{\frac{\xi}{0.1}}\sqrt{a_2}.
\ee
The parameter $a_2$ is naturally expected to be of order 1. In order not to worsen the tuning of the model it is thus important to have light enough fermionic top partners. 

In this paper we presented the 5D implementation of a specific class of composite Higgs models with a (pseudo-)Goldstone boson Higgs from the $SO(5)/SO(4)$ coset. Our starting point is the observation, already present in Refs.~\cite{Pomarol:2012vn, Panico:2012vr}, that in order for $\xi$ to be a good measure of fine tuning and not to underestimate it, the Higgs potential must be the sum of at least two independent periodic functions of the Higgs field $h$ which are generated at the same order in the elementary-composite mixing expansion. This fact constrains the fermionic content of the model. The simplest way to satisfy this requirement is to couple the left-handed top quark to a symmetric representation of $SO(5)$, a {\bf{14}}, and the right-handed top quark to an $SO(5)$ singlet.

Models in 5D are expected to be more constrained and to face a more severe fine tuning problem with respect to more general composite Higgs models. The reason for this are Eqs.~\eqref{outlook:S} and \eqref{outlook:higgsmass} and the fact that the masses of the fermionic and bosonic resonances are predicted to be parametrically the same, fixed by the size $L$ of the extra dimension. This implies that, once $m_\psi \sim m_\rho$ is fixed to comply with the bound on $\hat S$, an additional suppression of either $\xi$ or $a_2$ is needed to obtain the right Higgs mass.

We discussed these issues in our model finding that the NDA estimate in \Eqr{outlook:higgsmass} may be too pessimistic. 
In our model this occurs because of an additional, fortuitous, overall suppression of the estimate in \Eqr{outlook:higgsmass} by a factor $\sim 0.4\div 0.5$. The existence of such a suppression implies a somewhat heavier spectrum of the top partners can be obtained without worsening the fine tuning. Our model is somewhat peculiar in this respect because the requirement of a heavy enough top implies that some of the $SU(2)_L$ representations, into which the ${\bf{14}}$ decomposes, have to be anomalously light.
The existence of light, exotic charged, top-partners is probably the most important prediction of the class of composite Higgs models we have been discussing. Their experimental search is to be considered one of the most important avenues to either verify or falsify the composite Higgs idea.

A thorough discussion of the phenomenology of the fermionic sector of our model is beyond the scope of the present work. We will briefly discuss the salient features.
The QCD pair production of fermionic color triplets (the top partners) is sizable, ranging (including NLO effects) from 330 (570)~{\textrm{fb}} for a 500~{\textrm{GeV}} top partner to 1.3 (3.3)~{\textrm{fb}} for a 1~{\textrm{TeV}} one, at the LHC for 7 (8)~{\textrm{TeV}} center of mass energy.
The experimental signatures of the top partners depend strongly on their charge.\footnote{The electric charge of the top partner has little relevance for the pair production. On the other hand its value is crucial to describe the single production in association to either a $b$ or a $t$ quark. As discussed in Ref.~\cite{DeSimone:2012ul} single production is the most relevant process for top partners of higher masses due to the lower kinematical threshold. Furthermore the presence of a forward jet in the final state provides an important experimental handle to reduce the impact of backgrounds.} The most interesting signals arise from the decay of the charge 5/3 ($\chi$) and 8/3 ($\Upsilon$) fermions. By charge conservation the decay chain of these exotic quarks will involve the emission of up to 3 (for the charge 8/3 top partner) same sign $W$s resulting in a final state involving up to 6 $W$ bosons (plus at least 2 b quarks). This results in a sizable fraction of same-sign di-lepton and tri-lepton events.

In our model, and in all constructions where the Higgs is a PGB from $SO(5)/SO(4)$, the fermionic resonances fill degenerate $SO(4)$ multiplets before EWSB and in the absence of elementary-composite mixing. In our model the lightest among these $SO(4)$ representations is a ${\bf 9}$ which decomposes under $SU(2)_L\times U(1)_Y$ as ${\bf 3}_{5/3}\oplus {\bf 3}_{2/3}\oplus {\bf 3}_{-1/3}$. This $SO(4)$ multiplet contains, in particular, two charge 5/3 and a charge 8/3 quarks.
A rough estimate of the current experimental bounds can be inferred from the analysis of Ref.~\cite{CMS-PAS-B2G-12-012} looking for same-sign di-leptons in a sample of 19.6\,fb$^{-1}$ from the 8\,TeV LHC run. In the analysis the SM is extended by the addition of a single charge 5/3 Dirac fermion decaying with unit brancing ratio into $W^+ t$. A 95\% C.L. lower bound of $770$\,GeV is set on the mass of the fermion. The actual bound will be stronger in our model. Neglecting the other fermionic states in the {\bf 9} and the other $SO(4)$ multiplets, we still have two charge 5/3 and one charge 8/3 quarks. The pair production of the charge 5/3 fermions will give rise to same-sign di-leptons with a branching ratio of 4.5\% and to tri-leptons with a branching ratio of 3\%.
In the case of the charge 8/3 quark it is necessary to keep in mind that the decay $\Upsilon\to\chi W^+$ cannot proceed through a on-shell $W$ because $\Upsilon$ and $\chi$ are exactly degenerate at tree level as they do not mix with the elementary states. Their splitting is induced by electroweak effects at 1-loop and is of order 200\,MeV \cite{Cirelli:2005br}. To understand the signatures arising form $\Upsilon$ pair production one has to compare the following three-body decay channels: $\Upsilon\to \pi^+\chi$ or $\Upsilon\to \bar \ell\nu\chi$ through an off-shell $W$ boson and $\Upsilon\to W^+W^+ t$ through an off-shell $\chi$. The two former decay widths are proportional to $G_F f_{\pi}^{2}\Delta M^3$ and $G_F\Delta M^5$ respectively, while the latter is proportional to $M$ and thus expected to be dominant. This implies that the typical final state arising from $\Upsilon$ pair production will contain three opposite sign $W$ boson pairs and two $b$ quarks. This final state will decay to same-sign di-leptons 6\% of the times and to tri-leptons 6.5\% of the times.\footnote{Notice that if it were one of the first two three body decays to dominate, the leptons coming from the decay of the off-shell $W$ would have been too soft to be observable and the $\Upsilon$ would have effectively behaved like its charge 5/3 partner.} Even with this naive approach we expect more than a threefold increase in di-lepton signal rates with respect to those assumed in Ref.~\cite{CMS-PAS-B2G-12-012} and a correspondingly stronger bound on the mass of the fermions.

Since a generic feature of our model is the existence of an $SO(4)$ multiplet of top partners, the {\bf 9}, which is parametrically lighter than the others, the appropriate way to describe its phenomenology is along the lines of Ref.~\cite{DeSimone:2012ul}. We are aware of a group working in this direction \cite{MatsedonskyiToAppear}.

An orthogonal experimental avenue to follow in generic composite Higgs models is the search for bosonic vector resonances, that is excited $W$, $Z$ bosons and excited gluons. For this we refer the reader to the existing literature \cite{Giudice:1024017, Falkowski:2011ua, Bini:2011zb}. \footnote{Some aspects of the vector resonances phenomenology are also under investigation in forthcoming papers \cite{CLICToAppear,PappadopuloToAppear}.}

Finally, as in every composite Higgs model, the Higgs couplings to SM gauge bosons and fermions are modified. From this point of view our model is indistinguishable from other known composite Higgs models as the one described in \cite{Agashe:2005p372}. Deviations from the SM are controlled by a single parameter, $\xi$. Bounds on $\xi$ coming from Higgs coupling measurements are currently mild, requiring $\xi<0.3\div 0.4$ \cite{Falkowski:2013dza}.


\section*{Acknowledgments}
We would like to thank Ennio Salvioni and Javi Serra for collaboration in the initial stage of this project and Riccardo Barbieri, Roberto Contino and Christophe Grojean for discussions. We are especially grateful to Riccardo Rattazzi and Andrea Wulzer for many useful discussions and comments on the draft. The work of D.~P.~ is supported by the NSF Grant PHY-0855653.
A.~T.~has been partially supported by a Marie Curie Early Initial Training Network Fellowship of the European Community's Seventh Framework Programme under contract number PITN-GA-2008-237920-UNILHC and also by the Swiss National Science Foundation under contract 200020-138131. The work of R.~T.~was partly supported by the Spanish MICINN under grants CPAN CSD2007-00042 (Consolider-Ingenio 2010 Programme) and FPA2010-17747, by the Community of Madrid under grant HEPHACOS S2009/ESP-1473, by the Research Executive Agency (REA) of the European Union under the Grant Agreement number PITN-GA-2010-264564 (LHCPhenoNet) and by the ERC Advanced Grant no.~267985, Electroweak Symmetry Breaking, Flavour and Dark Matter: One Solution for Three Mysteries (DaMeSyFla). We finally thank the grant SNF Sinergia n.~CRSII2-141847.

\appendix

\section{$SO(5)$ generators}\l{SO5generators}
A basis for the generators in the fundamental representation of the $SO(5)$ algebra is given by
\begin{align}
& (T^{a}_{R})_{IJ} = \frac{i}{2} \left[ \frac{1}{2} \epsilon^{abc} \left(  \delta^b_I \delta^c_J - \delta^b_J \delta^c_I \right) + \left(  \delta^a_I \delta^4_J - \delta^a_J \delta^4_I \right) \right] (-1)^{\delta^a_{2}},\nonumber \\
& (T^{a}_{L})_{IJ} = \frac{i}{2} \left[ \frac{1}{2} \epsilon^{abc} \left(  \delta^b_I \delta^c_J - \delta^b_J \delta^c_I \right) - \left(  \delta^a_I \delta^4_J - \delta^a_J \delta^4_I \right) \right] (-1)^{\delta^a_{1}} ,\nonumber \\
& (T^{\hat{a}}_\uno)_{IJ} = - \frac{i}{\sqrt{2}} \left(  \delta^{\hat{a}}_I \delta^5_J - \delta^{\hat{a}}_J \delta^5_I \right),
\label{gen}
\end{align}
where $I, J = 1,\,\ldots,\, 5$, while $\hat{a} = 1,\,\ldots,\, 4$ and $a = 1, 2, 3$. 
The generators  $\{T^{a}_{R,L}\}$ and $\{T^{\hat{a}}_\uno\}$ represent the $SO(4) \cong SU(2)_L \times SU(2)_R$ and the $SO(5)/SO(4)$ subspaces of $SO(5)$ respectively.


\section{5D implementation in AdS$_5$ space}\l{App5D}
In this Appendix we describe the implementation of the MCHM$_{14}$ in AdS$_5$ space and illustrate some details about the flat space construction which were not discussed in the text. We consider a 5D space-time with metric
\be\l{geom1}
ds^{2}=a\(z\)^{2}\(\eta_{\mu\nu}dx^{\mu}dx^{\nu}-dz^{2}\)\,,
\ee
where the coordinate $z$ varies on the interval $[z^{\text{UV}},z^{\text{IR}}]$ and the warp function $a\(z\)$ is a regular and positive function satisfying $a(z_{UV})=1$. This parametrization includes flat and AdS$_5$ spaces with
\be
\bry{rlll}
\dst {\textrm{flat}}: &\dst  \quad a(z)=1, &\dst  \quad z_{UV}=0, &\dst  \quad z_{IR}\equiv L\,, \vspace{2mm} \\
\dst {\textrm{AdS}_5}: &\dst \quad a(z)=\frac{1}{kz}, &\dst  \quad z_{UV}=\frac{1}{k}, &\dst  \quad z_{IR}>z_{UV}\,.
\ery
\ee


\subsection{Gauge degrees of freedom}
The Lagrangian for the gauge fields in terms of the general metric of \Eqr{geom1} is given by
\be\l{gauge5D1ads}
S_{5D}^g=\int d^{4}x\int_{z_{UV}}^{z_{IR}} dz \sqrt{g} \left[ \frac{1}{4 g_{5}^{2}L_0} \Tr[F_{MN}^2] + \frac{1}{4g_{X}^{2}L_0}\left(F^{X}_{MN}\right)^2 \right] \, ,
\ee
where $L_0$ has the dimensions of length and is equal to $L$ in flat space and to $1/k$ in AdS$_5$.
We also add a UV localized term exactly as in \Eqr{gauge5D2}. To derive the effective Lagrangian in AdS$_5$, we can follow the same procedure as discussed in Section~\ref{App5Dgauge}. The whole difference with respect to flat space is encoded in the two form factors $\Pi_V$ and $\hat\Pi_V$ which read in AdS$_5$
\begin{eqnarray}\label{gaugeads}
\Pi_V(p)&=&p\frac{J_0(p z_{UV})Y_0(p z_{IR})-J_0(p z_{IR})Y_0(p z_{UV})}{J_1(p z_{UV})Y_0(p z_{IR})-J_0(p z_{IR})Y_1(p z_{UV})},\\
\hat\Pi_V(p)&=&p\frac{J_1(p z_{IR})Y_0(p z_{UV})-J_0(p z_{UV})Y_1(p z_{IR})}{J_1(p z_{IR})Y_1(p z_{UV})-J_1(p z_{UV})Y_1(p z_{IR})}\,.
\end{eqnarray}
The Higgs decay constant in this case is given by
\be
f^2=\frac{4}{g^2_5}\frac{1}{z_{IR}^2-z_{UV}^2}\approx\frac{4}{g^2_5}\frac{1}{z^2_{IR}}\,, 
\ee
where, in the last step, we assumed $z_{IR}\gg z_{UV}$. In the same limit the SM gauge couplings can be written as
\be\label{gaugecouplingads}
\bry{lll}
\dst \frac{1}{g^2}&=&\dst \frac{1}{g_2^2}+\frac{1}{g_5^2}\left( \log\frac{z_{IR}}{z_{UV}}- \frac{3}{8} s_h^2\right)\approx\frac{1}{g_2^2}+\frac{1}{g_5^2}\log\frac{z_{IR}}{z_{UV}},\vspace{2mm}\\ 
\dst \frac{1}{g^{'2}}&=&\dst \frac{1}{g_1^2}+\left(\frac{1}{g_5^2}+\frac{1}{g_X^2}\right)\left(\log\frac{z_{IR}}{z_{UV}}-\tfrac{3}{8} s_h^2\frac{g_X^2}{g_5^2+g_X^2}\right)\approx\frac{1}{g_1^2}+\left(\frac{1}{g_5^2}+\frac{1}{g_X^2}\right)\log\frac{z_{IR}}{z_{UV}}\, ,
\ery
\ee
where we approximated $\log \frac{z_{IR}}{z_{UV}} \gg 1$. In the same limit the $W$ boson mass is given by
\be
m_W^2=\frac{s_h^2}{z_{IR}^2\log\frac{z_{IR}}{z_{UV}}}.
\ee
The KK mass scale (looking for instance at the $W$ boson tower) is
\be
M_{KK}\approx\frac{2.4}{z_{IR}}+O\left(\frac{1}{z_{IR}\warp}\right).
\ee
and the $\hat S$ parameter is
\be
\hat S\approx\frac{3}{8}\frac{s_h^2}{\warp}.
\ee
Note that the inclusion of UV localized kinetic terms for the gauge fields is not strictly necessary to realize partial compositeness in AdS$_5$ space. The value of $g_2$ and $g_1$ may be interpreted as the strength of the gauge coupling at the scale $k$ \cite{Agashe:2002fu}: when the localized kinetic terms vanish, the gauge fields are strongly coupled at that scale.


\subsection{Fermionic degrees of freedom}\l{App5Dfermions}
Here we discuss various details about the fermionic sector of the model both in AdS$_5$ and flat space.
The expression of $\Psi_q$ in terms of eigenstates of the SM quantum numbers is
\small\be\l{complete14}
\Psi_{q}=\(\bry{ccccc}
 \f{1}{2}\(\zeta +i \Upsilon - \mathfrak{T}\)  & \f{1}{2} \(i\zeta +\Upsilon\)  & \f{1}{\sqrt{2}}\(\mathcal{B}+\mathcal{X}\) & \f{i}{\sqrt{2}} \(\mathcal{B}^{*}+\mathcal{X}^{*}\) & \f{1}{2}\(b^{(1)}+\chi^{(1)}\)  \\
 \f{1}{2}\(i \zeta +\Upsilon\)  & -\f{1}{2}\(\zeta +i \Upsilon + \mathfrak{T}\)  & \f{i}{\sqrt{2}}\(\mathcal{B}-\mathcal{X}\) & -\f{1}{\sqrt{2}}\(\mathcal{B}^{*}-\mathcal{X}^{*}\) &  \f{i}{2}\(b^{(1)}-\chi^{(1)}\)  \\
\f{1}{\sqrt{2}}\(\mathcal{B}+\mathcal{X}\) & \f{i}{\sqrt{2}}\(\mathcal{B}-\mathcal{X}\) & \f{1}{2} \ovl{\mathfrak{T}} +\ovl{\mathcal{T}} & i\mathcal{T} & \f{1}{2}\(t^{(1)}+i t^{(2)}\) \\
 \f{i}{\sqrt{2}} \(\mathcal{B}^{*}+\mathcal{X}^{*}\) & -\f{1}{\sqrt{2}}\(\mathcal{B}^{*}-\mathcal{X}^{*}\) & i \mathcal{T} &\f{1}{2} \ovl{\mathfrak{T}} -\ovl{\mathcal{T}}& \f{1}{2}\(it^{(1)}+t^{(2)}\) \\
 \f{1}{2}\(b^{(1)}+\chi^{(1)}\) & \f{i}{2} \(b^{(1)}-\chi^{(1)}\) & \f{1}{2}\(t^{(1)}+i t^{(2)}\) & \f{1}{2}\(it^{(1)}-t^{(2)}\) & \f{2}{\sqrt{5}} t^{(3)} 
   \ery\)\,.
\ee\normalsize
where the fields $t^{(i)}$, $b^{(i)}$, $\chi^{(i)}$, $\zeta$ and $\Upsilon$ have charges $2/3$, $-1/3$, $5/3$, $-4/3$, $8/3$ respectively and
\be
\bry{lll}
\mathcal{B}^{(*)} & = &\f{1}{2}\(b^{(2)}+^{(-)}i b^{(3)}\)\,,\\
\mathcal{X}^{(*)} & = &\f{1}{2}\(\chi^{(2)}+^{(-)}i \chi^{(3)}\)\,,\\
\mathcal{T} & = &\f{1}{2}\(t^{(5)}+t^{(6)}\)\,,\\
\ovl{\mathcal{T}} & = &\f{1}{2}\(t^{(5)}-t^{(6)}\)\,,\\
\mathfrak{T} & = & t^{(4)} + \frac{1}{\sqrt{5}} t^{(3)}\,,\\
\ovl{\mathfrak{T}} & = & t^{(4)} - \frac{1}{\sqrt{5}} t^{(3)}\,.\\
\ery
\ee
The total 5D action for the fermionic sector of the model is given by
\be\l{5Dtotalaction}
S_{\text{total}}^{f}= S_{5D}^{f}+S_{\text{IR}}^{f}+\Delta S_{\text{UV}}+\Delta S_{\text{IR}} \, ,
\ee
where
\begin{align}
& \bry{llll}
\dst S_{5D}^{f}&=&\dst \int d^{4}x\int_{z_{UV}}^{Z_{IR}} \frac{dz}{L_0}\sqrt{g}~ \Bigg[ &
\dst \f{i}{2}\Tr \left[ \ovl{\Psi}^{q}e^M_A\Gamma^{A}D_{M}\Psi^{q}-(D_{M}\Psi^{q\dagger})e^M_A\Gamma^0\Gamma^{A}\Psi^{q} \right]-M_{\Psi_{Q}}\ovl{\Psi}^{q}\Psi^{q}\vspace{2mm}\\
&&&\dst +\f{i}{2}\Bigg(\ovl{\Psi}^{u}e^M_A\Gamma^{A}D_{M}\Psi^{u}-(D_{M}\Psi^{u\dagger})e^M_A\Gamma^0\Gamma^{A}\Psi^{u}\Bigg)-M_{\Psi_{u}}\ovl{\Psi}^{u}\Psi^{u}\Bigg]\,,
\ery\\
& \bry{llll}
\dst S_{\text{IR}}^{f}&=&\dst \int d^{4}x~\sqrt{g_{IR}}~ \bigg[ & \left( m_{11} \ovl{\psi}^{\(\mathbf{1}\)}_{qR} \psi_{tL}+\textrm{h.c.}\right)+\left(i k^t_1\,\ovl{\psi}_{tL} e^\mu_a\gamma^a\partial_\mu \psi_{tL}+ik^q_1\,\ovl{\psi}^{\(\mathbf{1}\)}_{qR}e^\mu_a\gamma^a\partial_\mu \psi^{\(\mathbf{1}\)}_{qR} \right. \vspace{2mm} \\ 
&&&\dst \left.+~ik^{q}_{4}\,\ovl{\psi}^{\(\mathbf{4}\)}_{qL}e^\mu_a\gamma^a\partial_\mu \psi^{\(\mathbf{4}\)}_{qL}+ik^{q}_{9}\,\ovl{\psi}^{\(\mathbf{9}\)}_{qL}e^\mu_a\gamma^a\partial_\mu \psi^{\(\mathbf{9}\)}_{qL}\right) \bigg]\,,
\ery\\
& \bry{lll} 
\dst  \Delta S_{\text{UV}}&=&\dst  \f{1}{L_0}\int d^{4}x~\sqrt{g_{UV}} ~\frac{1}{2}\left[ \ovl{\Psi}_{q}\Psi_{q}-\ovl{\Psi}_{t}\Psi_{t}\right]\,,\ery\vspace{2mm}\\
& \bry{lll} 
\dst \Delta S_{\text{IR}}&=&\dst \f{1}{L_0}\int d^{4}x~\sqrt{g_{IR}}~\frac{1}{2}\Big[\ovl{\psi}^{\(\mathbf{1}\)}_{q}\psi^{\(\mathbf{1}\)}_{q}-\ovl{\psi}^{\(\mathbf{4}\)}_{q}\psi^{\(\mathbf{4}\)}_{q}-\ovl{\psi}^{\(\mathbf{9}\)}_{q}\psi^{\(\mathbf{9}\)}_{q}-\ovl{\Psi}^{u}\Psi^{u}\Big]\,.\ery
\end{align}
$L_0$ is a length we use to normalize the bulk action in order to have 5D fermions with the same mass dimension as 4D fermions. As in the case of the gauge action it is equal to $L$ in flat space and to $1/k$ in AdS$_5$. $g_{IR}$ and $g_{UV}$ are the induced metrics on the IR and UV boundaries respectively. $e^M_A$ is the f\"{u}nfbein and $e^\mu_a$ its projection on the 4D indices. It is given by $\delta^{M}_A/a(z)$ for flat and AdS$_5$ metric. Covariant derivatives in $S_{5D}^f$ include both gauge and spin connections. The latter cancel out in the equations of motion.
In flat space we also add localized kinetic terms on the UV boundary, as in \Eqr{UVfermions}, to properly implement partial compositeness.

The bulk equations of motion corresponding to \Eqr{5Dtotalaction} are
\be\l{eom1}
\left[ \de_{5} +2\f{\de_{5} a\(z\)}{a\(z\)}\pm M_{\Psi^{i}}a\(z\) \right]\Psi^{i}_{L,R}=\pm \psl\Psi^{i}_{R,L}\,.
\ee
These are supplemented by the following boundary conditions:
\bit
\item {\bf UV boundary conditions}
\be
\Psi_{qL}\(z_{UV}\)=U\Psi_{qL}^{0}U^{T}\qquad\qquad \delta\Psi_{qL}\(z_{UV}\)=\delta\Psi_{qL}^{0}=0\,,
\ee
with $\Psi_{qL}^{0}$ given by \Eqr{complete14} where all the fields but $t^{1}=t_{L}$ and $b^{1}=b_{L}$ are set to zero while $\Psi_{qR}\(z_{UV}\)$ is free to vary and 
\be
\Psi_{tR}\(z_{UV}\)=\Psi_{tR}^{0}\qquad\qquad \delta\Psi_{tR}\(z_{UV}\)=\delta\Psi_{tR}^{0}=0\,,
\ee
with $\Psi_{tR}^{0}=t_{R}$ and $\Psi_{tL}\(z_{UV}\)$ free to vary.
\item {\bf IR boundary conditions}
\begin{align}
& \dst \Psi_{tR}\(z_{IR}\)=L_0 m_{11}\, \psi^{\(\mathbf{1}\)}_{qR}\(z_{IR}\)\,, \qquad && \dst \Psi_{tL}\(z_{IR}\)=-\f{1}{L_0 m_{11} }\,\psi^{\(\mathbf{1}\)}_{qL}\(z_{IR}\)\,,\\
& \dst \psi^{\(\mathbf{4}\)}_{qR}\(z_{IR}\)= k_{4}^{q} \,\frac{L_0}{a(z_{IR})}\psl\, \psi^{\(\mathbf{4}\)}_{qL}\(z_{IR}\) \quad && \dst \psi^{\(\mathbf{9}\)}_{qR}\(z_{IR}\)=k_{9}^{q} \,\frac{L_0}{a(z_{IR})}\psl\,\psi^{\(\mathbf{9}\)}_{qL}\(z_{IR}\)\,.
\end{align}
where, without loss of generality, we have set $k_{1}^{q}=k_{1}^{t}=0$.
\eit
The $U$ transformation in the $\Psi_{qL}$ UV boundary condition has been introduced to take the $SO(5)$ transformation in \Eqr{holographicgauge1} into account, needed to rotate the Higgs ($A^{5}$) away from the bulk and the IR boundary. The SM quantum numbers and the boundary conditions for all fermionic fields of the 5D model are summarized in Table \ref{Table1}.

Solving the bulk equations of motion \eqref{eom1} with the boundary conditions above, integrating out the bulk fields and matching to the 4D effective action \eqref{efflag14}  gives the form factors as functions of the fermion bulk-to-boundary propagators $G_{+}\(z,m\)$ and $G_{-}\(z,m\)$, which are functions of the metric $g_{\mu\nu}$. 
Defining $\omega_{q, t}^{2}=p^{2}-m_{q, t}^{2}$ in flat space and \sloppy\mbox{$\omega_{q} = 2/ (\pi z_{UV})$} and $\omega_{t} = 2/ (\pi z_{IR})$ in AdS$_5$ the form factors can be written in the form
\bes\small \l{FFAdS}
\begin{align}
& \Pi_{0}^{Q}=\f{G^{+}_{R}(z_{UV}, M_{\Psi_{q}})-k_{9}^{q}\,pL_{0}\,G^{-}_{R}(z_{UV}, M_{\Psi_{q}})}{pL_{0} \,\big[G^{+}_{L}(z_{UV}, M_{\Psi_{q}})-k_{9}^{q}\,pL_{0}\,G^{-}_{L}(z_{UV}, M_{\Psi_{q}})\big]}\,, \\
& \tilde{\Pi}_{0}^{Q}=Z_{q}+\Pi_{0}^{Q}\,, \\
& \Pi_{1}^{Q}=\f{\omega_{q}^{2}\(k_{4}^{q}\,-k_{9}^{q}~\,\)}{2 p^{2} \big[G^{+}_{L}(z_{UV}, M_{\Psi_{q}})-k_{4}^{q}\,pL_{0}\,G^{-}_{L}(z_{UV}, M_{\Psi_{q}})\big]\big[G^{+}_{L}(z_{UV}, M_{\Psi_{q}})-k_{9}^{q}\,pL_{0}\,G^{-}_{L}(z_{UV}, M_{\Psi_{q}})\big]}\,,\\
& \bry{lll} \dst \Pi_{2}^{Q}&=&\dst \frac{\omega_{q}^{2}}{4 p^{2} G^{+}_{L}(M_{\Psi_{q}})}\Bigg[\f{3k_{9}^{q}}{G^{+}_{L}(z_{UV}, M_{\Psi_{q}})-k_{9}^{q}\,pL_{0}\,G^{-}_{L}(z_{UV}, M_{\Psi_{q}})}\\
&&\dst -\f{8 k_{4}^{q}}{G^{+}_{L}(z_{UV}, M_{\Psi_{q}})-k_{4}^{q}\,pL_{0}\,G^{-}_{L}(z_{UV}, M_{\Psi_{q}})}\\
&&\dst -\f{5G^{+}_{R}(z_{UV}, M_{\Psi_{t}})}{p^{2} L_{0}\,\big[G^{-}_{L}(z_{UV}, M_{\Psi_{q}}) G^{+}_{R}(z_{UV}, M_{\Psi_{t}}) + m_{11}^{2}L_{0}^{2}\,G^{+}_{L}(z_{UV}, M_{\Psi_{q}}) G^{-}_{R}(z_{UV}, M_{\Psi_{t}})\big]}\Bigg]\,,\ery \\
& \Pi_{0}^{t}=-\f{G^{-}_{L}(z_{UV}, M_{\Psi_{q}}) G^{+}_{L}(z_{UV}, M_{\Psi_{t}})+m_{11}^{2}L_{0}^{2}\,G^{-}_{L}(z_{UV}, M_{\Psi_{t}}) G^{+}_{L}(z_{UV}, M_{\Psi_{q}})}{pL_{0}\,\big[G^{-}_{L}(z_{UV}, M_{\Psi_{q}}) G^{+}_{R}(z_{UV}, M_{\Psi_{t}}) + m_{11}^{2}L_{0}^{2}\,G^{+}_{L}(z_{UV}, M_{\Psi_{q}}) G^{-}_{R}(z_{UV}, M_{\Psi_{t}})\big]}\,, \\
& \tilde{\Pi}_{0}^{t}=Z_{t}+\Pi_{0}^{t}\,,\\
& M_{1}^{t}= \frac{- i\sqrt{5} m_{11}\omega_{q}\omega_{t}}{2 p^{2} \big[L_{0}^{2}m_{11}^{2}\,G^{+}_{L}(z_{UV}, M_{\Psi_{q}}) G^{-}_{R}(z_{UV}, M_{\Psi_{t}})+G^{-}_{L}(z_{UV}, M_{\Psi_{q}}) G^{+}_{R}(z_{UV}, M_{\Psi_{t}})\big]}\,.
\end{align}
\ees
The bulk-to-boundary propagators $G^{\pm}_{L}\(z,m\)$ and $G^{\pm}_{R}\(z,m\)$ are given by the following expressions:
\bit
\item Flat space ($z_{IR}=L$)
\be
\bry{l}
\dst G^{+}_{L}\(z,m\)= \f{1}{p} \( \omega\cos \omega \(L-z\)+m \sin\omega\(L-z\) \) \,,\vspace{2mm} \\
\dst G^{-}_{L}\(z,m\)= \sin\omega\(L-z\)\,, \vspace{2mm}\\
\dst G^{+}_{R}\(z,m\)= \sin\omega\(L-z\)\,, \vspace{2mm}\\
\dst G^{-}_{R}\(z,m\)=- \f{1}{p} \( \omega \cos \omega \(L-z\) + m \sin\omega\(L-z\) \)\,.
\ery
\ee
\item AdS$_{5}$
\be
\bry{l}
\dst G^{\pm}_{L}\(z,m\)= Y_{ML \mp 1/2}\(pz_{IR}\)J_{ML+1/2}\(pz\)-J_{ML \mp 1/2}\(pz_{IR}\)Y_{ML+1/2}\(pz\) \\
\dst G^{\pm}_{R}\(z,m\)= Y_{ML \mp 1/2}\(pz_{IR}\)J_{ML-1/2}\(pz\)-J_{ML \mp 1/2}\(pz_{IR}\)Y_{ML-1/2}\(pz\) \,.
\ery
\ee
As the propagators appear only as functions of $z_{UV}$ in \Eqr{FFAdS}, for clarity we have absorbed a factor of $z^{5/2}$,   originating from the $G$, inside $\omega_{q, t}$.
\eit

\begin{table}[t]\label{Table1}
\begin{center}
\begin{tabular}{cccccccc}
&$T_{3}^{L}$ &$T_{3}^{R}$&$X$&$Y$&$Q$& $\psi_{L}$ BC (UV,IR)&$\psi_{R}$ BC (UV,IR)\\ \hline
\red{$t^{(1)}$}& \red{$+\f{1}{2}$} & \red{$-\f{1}{2}$} & \red{$+\f{2}{3}$} & \red{$+\f{1}{6}$} & \red{$+\f{2}{3}$} & \red{$(t_{L},/)$} & \red{$(/, k_{4}^{q}\, L_{0} \psl\, t_{L}^{(1)})$} \\
$t^{(2)}$& $-\f{1}{2}$ & $+\f{1}{2}$ & $+\f{2}{3}$ & $+\f{7}{6}$ & $+\f{2}{3}$ & $(0,/)$ & $(/, k_{4}^{q}\,L_{0} \psl\,t_{L}^{(2)})$ \\
\blue{$t^{(3)}$}& \blue{$0$} & \blue{$0$} & \blue{$+\f{2}{3}$} & \blue{$+\f{2}{3}$} & \blue{$+\f{2}{3}$} & \blue{$(0,-\tfrac{2}{\sqrt{5}}L_{0} m_{11} t^{(7)}_{L})$} & \blue{$(/,t^{(3)}_{R})$} \\
\blue{$t^{(4)}$}& \blue{$0$} & \blue{$0$} & \blue{$+\f{2}{3}$} & \blue{$+\f{2}{3}$} & \blue{$+\f{2}{3}$} & \blue{$(0,/)$} & \blue{$(/, k_{9}^{q}\, L_{0}\psl\,t_{L}^{(4)})$} \\
$t^{(5)}$& $+1$ & $-1$ & $+\f{2}{3}$ & $-\f{1}{3}$ & $+\f{2}{3}$ & $(0,/)$ & $(/, k_{9}^{q}\,L_{0}\psl\,t_{L}^{(5)})$ \\
$t^{(6)}$& $-1$ & $1$ & $+\f{2}{3}$ & $+\f{5}{3}$ & $+\f{2}{3}$ & $(0,/)$ & $(/, k_{9}^{q}\,L_{0}\psl\,t_{L}^{(6)})$ \\
\red{$b^{(1)}$}& \red{$-\f{1}{2}$} & \red{$-\f{1}{2}$} & \red{$+\f{2}{3}$} & \red{$+\f{1}{6}$} & \red{$-\f{1}{3}$} & \red{$(b_{L},/)$} & \red{$(/,k_{4}^{q}\,L\psl\,b_{L}^{(1)})$} \\
$b^{(2)}$& $0$ & $-1$ & $+\f{2}{3}$ & $-\f{1}{3}$ & $-\f{1}{3}$ & $(0,/)$ & $(/, k_{9}^{q}\,L_{0}\psl\,b_{L}^{(2)})$ \\
$b^{(3)}$& $-1$ & $0$ & $+\f{2}{3}$ & $+\f{2}{3}$ & $-\f{1}{3}$ & $(0,/)$ & $(/, k_{9}^{q}\,L_{0}\psl\,b_{L}^{(3)})$ \\
\grey{$\chi^{(1)}$ } & \grey{$+\f{1}{2}$} & \grey{$+\f{1}{2}$} & \grey{$+\f{2}{3}$} & \grey{$+\f{7}{6}$} & \grey{$+\f{5}{3}$} & \grey{$(0,/)$} & \grey{$(/, k_{4}^{q}\,L_{0} \psl\,\chi^{(1)}_{L})$} \\
\grey{ $\chi^{(2)}$ } & \grey{$+1$} & \grey{$0$} & \grey{$+\f{2}{3}$} & \grey{$+\f{2}{3}$} & \grey{$+\f{5}{3}$} & \grey{$(0,/)$} & \grey{$(/, k_{9}^{q}\,L_{0}\psl\,\chi^{(2)}_{L})$} \\
\grey{ $\chi^{(3)}$ } & \grey{$0$} & \grey{$+1$} & \grey{$+\f{2}{3}$} & \grey{$+\f{5}{3}$} & \grey{$+\f{5}{3}$} & \grey{$(0,/)$} & \grey{$(/, k_{9}^{q}\,L_{0}\psl\,\chi^{(3)}_{L})$} \\
\grey{ $\zeta$ }& \grey{$-1$} & \grey{$-1$} & \grey{$+\f{2}{3}$} & \grey{$-\f{1}{3}$} & \grey{$-\f{4}{3}$} & \grey{$(0,/)$} & \grey{$(/, k_{9}^{q}\,L_{0} \psl\,\zeta_{L})$} \\
\grey{ $\Upsilon$ } & \grey{$+1$} & \grey{$+1$} & \grey{$+\f{2}{3}$} & \grey{$+\f{5}{3}$} & \grey{$+\f{8}{3}$} & \grey{$(0,/)$} & \grey{$(/, k_{9}^{q}\,L_{0}\psl\,\Upsilon_{L})$} \\
\hline
\blue{$\Psi^{t}\equiv t^{(7)}$}& \blue{$0$} & \blue{$0$} & \blue{$+\f{2}{3}$} & \blue{$+\f{2}{3}$} & \blue{$+\f{2}{3}$} & \blue{$(/,t_{L}^{(7)})$} & \blue{$(t_{R},\tfrac{\sqrt{5}}{2}L_{0} m_{11}\,t^{(3)}_{R})$} \\
\end{tabular}
\end{center}
\caption{Quantum numbers of the 5D fermion fields of the model. In the last row the boundary conditions on the UV and IR branes are shown. In \red{red} (\blue{blue}) we indicate the states with the same SM quantum numbers of the \red{$q_{L}$} and \blue{$t_{R}$}. The holographic fields are $t_{L}$, $b_{L}$ and $t_{R}$.}
\end{table}

%


\section{The electroweak fit}\label{ewfit}
We use the following experimental determination of the $\hat S$ and $\hat T$ parameters
\be
\begin{array}{ccc}
\hat S &=&(0.39\pm0.70)\,10^{-3}\\
\hat T &=&(0.60\pm0.56)\,10^{-3}
\end{array}
\ee
with the corresponding correlation matrix
\be
\rho=\begin{pmatrix}
1& 0.91\\
0.91&1
\end{pmatrix}.
\ee
The resulting $\chi^2$ function is given by
\be\nonumber
\chi^2(\hat S, \hat T)= V_i (\sigma^{2})^{-1}_{ij} V_j,
\ee
\be
V=(\hat S-\hat S_0,\hat T-\hat T_0),\quad \sigma^{2}_{ij}=\sigma_i\rho_{ij}\sigma_j.
\ee
The contributions to $\hat S$ and $\hat T$ in our model are estimated by
\begin{eqnarray}\label{ewo}
\hat S&=&\hat S_{UV}+\frac{\alpha_{EM}(m_Z)}{48\pi s_W^2}\xi\log\left(\frac{\Lambda^2}{m_h^2}\right) \, ,\\
\hat T&=& -\frac{3\alpha_{EM}(m_Z)}{16\pi c_W^2}\xi\log\left(\frac{\Lambda^2}{m_h^2}\right) \, , 
\end{eqnarray}
where $\Lambda$ is the scale at which the logs are cut off. We fix $\Lambda$ to be the mass of the first vector KK resonance. $\hat S$ and $\hat T$ are defined to vanish for a SM Higgs boson with $m_h=125\,{\textrm{GeV}}$.
%

\bibliographystyle{mine}
\bibliography{bibliography}

\providecommand{\href}[2]{#2}\begingroup\raggedright\begin{thebibliography}{10}

\bibitem{Contino:2003p378}
R.~Contino, Y.~Nomura and A.~Pomarol, ``{Higgs as a Holographic
  Pseudo-Goldstone Boson}'',
  \href{http://dx.doi.org/10.1016/j.nuclphysb.2003.08.027}{{\em Nucl. Phys.}
  {\bfseries B 671} (2003) 148--174},
  \href{http://xxx.lanl.gov/abs/hep-ph/0306259}{{\tt hep-ph/0306259}} [\href{http://inspirehep.net/record/622164}{Inspire}].

\bibitem{Agashe:2004ib}
K.~Agashe, R.~Contino and A.~Pomarol, ``The {M}inimal {C}omposite {H}iggs
  {M}odel'', \href{http://dx.doi.org/10.1016/j.nuclphysb.2005.04.035}{{\em
  Nucl. Phys.} {\bfseries B 719} (2005) 165--187},
  \href{http://xxx.lanl.gov/abs/hep-ph/0412089}{{\tt hep-ph/0412089}} [\href{http://inspirebeta.net/record/666275}{Inspire}].

\bibitem{Agashe:2005p372}
K.~Agashe and R.~Contino, ``{The Minimal Composite Higgs Model and Electroweak
  Precision Tests}'',
  \href{http://dx.doi.org/10.1016/j.nuclphysb.2006.02.011}{{\em Nucl. Phys.}
  {\bfseries B 742} (2006) 59--85},
  \href{http://xxx.lanl.gov/abs/hep-ph/0510164}{{\tt hep-ph/0510164}} [\href{http://inspirehep.net/record/694895}{Inspire}].

\bibitem{Contino:2006fd}
R.~Contino, L.~{Da Rold} and A.~Pomarol, ``Light custodians in natural
  composite {H}iggs models'',
  \href{http://dx.doi.org/10.1103/PhysRevD.75.055014}{{\em Phys. Rev.}
  {\bfseries D 75} (2007) 055014},
  \href{http://xxx.lanl.gov/abs/hep-ph/0612048}{{\tt hep-ph/0612048}} [\href{http://inspirebeta.net/record/733499}{Inspire}].

\bibitem{Panico:2005kd}
G.~Panico, M.~Serone and A.~Wulzer, ``{A Model of Electroweak Symmetry
  Breaking from a Fifth Dimension}'',
  \href{http://dx.doi.org/10.1016/j.nuclphysb.2006.01.025}{{\em Nucl. Phys.}
  {\bfseries B739} (2006) 186--207},
  \href{http://xxx.lanl.gov/abs/hep-ph/0510373}{{\tt hep-ph/0510373}} [\href{http://inspirehep.net/record/696326}{Inspire}].

\bibitem{Barbieri:2007bh}
R.~Barbieri, B.~Bellazzini, S.~Rychkov and A.~Varagnolo, ``{The Higgs boson
  from an extended symmetry}'',
  \href{http://dx.doi.org/10.1103/PhysRevD.76.115008}{{\em Phys. Rev.}
  {\bfseries D76} (2007) }, \href{http://xxx.lanl.gov/abs/0706.0432}{{\tt
  arXiv:0706.0432}} [\href{http://inspirebeta.net/record/752225}{Inspire}].

\bibitem{Contino:2010mr}
R.~Contino, ``{Tasi 2009 lectures: The Higgs as a Composite Nambu-Goldstone
  Boson}'', \href{http://xxx.lanl.gov/abs/1005.4269}{{\tt arXiv:1005.4269}} [\href{http://inspirehep.net/record/856065}{Inspire}].

\bibitem{Barbieri:2012wia}
R.~Barbieri, D.~Buttazzo, F.~Sala, D.~Straub and A.~Tesi, ``{A 125 GeV
  composite Higgs boson versus flavour and electroweak precision tests}'',
  \href{http://xxx.lanl.gov/abs/1211.5085}{{\tt arXiv:1211.5085}} [\href{http://inspirehep.net/record/1203476}{Inspire}].

\bibitem{1991NuPhB.365..259K}
D.~B. Kaplan, ``{Flavor at SSC energies: A new mechanism for dynamically
  generated fermion masses}'',
  \href{http://dx.doi.org/10.1016/S0550-3213(05)80021-5}{{\em Nucl. Phys.}
  {\bfseries B 365} (1991) 259} [\href{http://inspirehep.net/record/314641}{Inspire}].

\bibitem{KerenZur:2012td}
B.~Keren-Zur, P.~Lodone, M.~Nardecchia, D.~Pappadopulo, R.~Rattazzi and
  L.~Vecchi, ``{On Partial Compositeness and the CP asymmetry in charm
  decays}'', \href{http://dx.doi.org/10.1016/j.nuclphysb.2012.10.012}{{\em
  Nucl.Phys.} {\bfseries B867} (2013) 429--447},
  \href{http://xxx.lanl.gov/abs/1205.5803}{{\tt arXiv:1205.5803}} [\href{http://inspirehep.net/record/81406}{Inspire}].

\bibitem{Panico:2012vr}
G.~Panico, M.~Redi, A.~Tesi and A.~Wulzer, ``{On the Tuning and the Mass of
  the Composite Higgs}'', \href{http://xxx.lanl.gov/abs/1210.7114}{{\tt
  arXiv:1210.7114}} [\href{http://inspirehep.net/record/1193785}{Inspire}].

\bibitem{Callan:1969p1799}
C.~G. Callan, S.~Coleman, J.~Wess and B.~Zumino, ``Structure of
  {P}henomenological {L}agrangians. {II}'',
  \href{http://dx.doi.org/10.1103/PhysRev.177.2247}{{\em Phys. Rev.} {\bfseries
  177} (1969) 2247--2250} [\href{http://inspirebeta.net/record/54960}{Inspire}].

\bibitem{Giudice:1024017}
G.~F. Giudice, C.~Grojean, A.~Pomarol and R.~Rattazzi, ``{The
  Strongly-Interacting Light Higgs. }'',
  \href{http://dx.doi.org/10.1088/1126-6708/2007/06/045}{{\em JHEP} {\bfseries
  06} (2007) 045}, \href{http://xxx.lanl.gov/abs/hep-ph/0703164}{{\tt
  hep-ph/0703164}} [\href{http://inspirehep.net/record/746568}{Inspire}].

\bibitem{Pomarol:2012vn}
A.~Pomarol and F.~Riva, ``{The Composite Higgs and Light Resonance
  Connection}'', \href{http://dx.doi.org/10.1007/JHEP08(2012)135}{{\em JHEP}
  {\bfseries 08} (2012) 135}, \href{http://xxx.lanl.gov/abs/1205.6434}{{\tt
  arXiv:1205.6434}} [\href{http://inspirehep.net/record/1116432}{Inspire}].

\bibitem{Marzocca:2012tt}
D.~Marzocca, M.~Serone and J.~Shu, ``{General Composite Higgs Models}'',
  \href{http://dx.doi.org/10.1007/JHEP08(2012)013}{{\em JHEP} {\bfseries 08}
  (2012) 013}, \href{http://xxx.lanl.gov/abs/1205.0770}{{\tt arXiv:1205.0770}} [\href{http://inspirehep.net/record/1113466}{Inspire}].

\bibitem{Agashe:2006at}
K.~Agashe, R.~Contino, L.~{Da Rold} and A.~Pomarol, ``{A Custodial symmetry
  for $Z b \bar{b}$}'',
  \href{http://dx.doi.org/10.1016/j.physletb.2006.08.005}{{\em Phys.Lett.}
  {\bfseries B641} (2006) }, \href{http://xxx.lanl.gov/abs/hep-ph/0605341}{{\tt
  hep-ph/0605341}} [\href{http://inspirehep.net/record/718059}{Inspire}].

\bibitem{Mrazek:2011iu}
J.~Mrazek, A.~Pomarol, R.~Rattazzi, M.~Redi, J.~Serra and A.~Wulzer, ``{The
  Other Natural Two Higgs Doublet Model}'',
  \href{http://dx.doi.org/10.1016/j.nuclphysb.2011.07.008}{{\em Nucl.Phys.}
  {\bfseries B 853} (2011) }, \href{http://xxx.lanl.gov/abs/1105.5403}{{\tt
  arXiv:1105.5403}} [\href{http://inspirehep.net/record/901804}{Inspire}].

\bibitem{Coleman:1973jx}
S.~R. Coleman and E.~J. Weinberg, ``{Radiative Corrections as the Origin of
  Spontaneous Symmetry Breaking}'',
  \href{http://dx.doi.org/10.1103/PhysRevD.7.1888}{{\em Phys. Rev.} {\bfseries
  D 7} (1973) 1888--1910} [\href{http://inspirehep.net/record/81406}{Inspire}].

\bibitem{Panico:2007fq}
G.~Panico and A.~Wulzer, ``{Effective Action and Holography in 5D Gauge
  Theories}'', \href{http://dx.doi.org/10.1088/1126-6708/2007/05/060}{{\em
  JHEP} {\bfseries 05} (2007) 060},
  \href{http://xxx.lanl.gov/abs/hep-th/0703287}{{\tt hep-th/0703287}} [\href{http://inspirehep.net/record/747668}{Inspire}].

\bibitem{Serone:2009ks}
M.~Serone, ``{Holographic Methods and Gauge-Higgs Unification in Flat Extra
  Dimensions}'', \href{http://dx.doi.org/10.1088/1367-2630/12/7/075013}{{\em
  New J. Phys.} {\bfseries 12} (2010) 075013},
  \href{http://xxx.lanl.gov/abs/0909.5619}{{\tt arXiv:0909.5619}} [\href{http://inspirehep.net/record/832650}{Inspire}].

\bibitem{Scrucca:2003cf}
C.~Scrucca, M.~Serone and L.~Silvestrini, ``{Electroweak symmetry breaking and
  fermion masses from extra dimensions}'',
  \href{http://dx.doi.org/10.1016/j.nuclphysb.2003.07.013}{{\em Nucl.Phys.}
  {\bfseries B669} (2003) 128--158},
  \href{http://xxx.lanl.gov/abs/hep-ph/0304220}{{\tt hep-ph/0304220}} [\href{http://inspirehep.net/record/617466}{Inspire}].

\bibitem{Barbieri:678689}
R.~Barbieri, A.~Pomarol and R.~Rattazzi, ``{Weakly coupled Higgsless theories
  and precision electroweak tests}'',
  \href{http://dx.doi.org/10.1016/j.physletb.2004.04.005}{{\em Physics Letters
  B} {\bfseries 591} (2003) 141--149},
  \href{http://xxx.lanl.gov/abs/hep-ph/0310285}{{\tt hep-ph/0310285}} [\href{http://inspirebeta.net/record/631485}{Inspire}].

\bibitem{Panico:2010ir}
G.~Panico, M.~Safari and M.~Serone, ``{Simple and Realistic Composite Higgs
  Models in Flat Extra Dimensions}'',
  \href{http://dx.doi.org/10.1007/JHEP02(2011)103}{{\em JHEP} {\bfseries 1102}
  (2011) 103}, \href{http://xxx.lanl.gov/abs/1012.2875}{{\tt arXiv:1012.2875}} [\href{http://inspirebeta.net/record/881254}{Inspire}].

\bibitem{Witten:1983ut}
E.~Witten, ``{Some Inequalities Among Hadron Masses}'',
  \href{http://dx.doi.org/10.1103/PhysRevLett.51.2351}{{\em Phys. Rev. Lett.}
  {\bfseries 51} (1983) 2351} [\href{http://inspirehep.net/record/193973}{Inspire}].

\bibitem{1984NuPhB.234..173V}
C.~Vafa and E.~Witten, ``{Restrictions on symmetry breaking in vector-like
  gauge theories}'', \href{http://dx.doi.org/10.1016/0550-3213(84)90230-X}{{\em
  Nucl. Phys.} {\bfseries B 234} no.~1, (1984) 173} [\href{http://inspirehep.net/record/14056}{Inspire}].

\bibitem{Contino:2004cj}
R.~Contino and A.~Pomarol, ``{Holography for fermions}'',
  \href{http://dx.doi.org/10.1088/1126-6708/2004/11/058}{{\em JHEP} {\bfseries
  11} (2004) 058}, \href{http://xxx.lanl.gov/abs/hep-th/0406257}{{\tt
  hep-th/0406257}} [\href{http://inspirehep.net/record/653279}{Inspire}].

\bibitem{Panico:1359049}
G.~Panico and A.~Wulzer, ``{The Discrete Composite Higgs Model}'',
  \href{http://dx.doi.org/10.1007/JHEP09(2011)135}{{\em JHEP} {\bfseries 1109}
  (2011) 135}, \href{http://xxx.lanl.gov/abs/1106.2719}{{\tt arXiv:1106.2719}} [\href{http://inspirehep.net/record/913588}{Inspire}].

\bibitem{Redi:qi}
S.~D. Curtis, M.~Redi and A.~Tesi, ``{The 4D Composite Higgs}'',
  \href{http://dx.doi.org/10.1007/JHEP04(2012)042}{{\em JHEP} {\bfseries 1204}
  (2012) 042}, \href{http://xxx.lanl.gov/abs/1110.1613}{{\tt arXiv:1110.1613}} [\href{http://inspirebeta.net/record/666275}{Inspire}].

\bibitem{Redi:2012vq}
M.~Redi and A.~Tesi, ``{Implications of a Light Higgs in Composite Models}'',
  \href{http://dx.doi.org/10.1007/JHEP10(2012)166}{{\em JHEP} {\bfseries 1210}
  (2012) 166}, \href{http://xxx.lanl.gov/abs/1205.0232}{{\tt arXiv:1205.0232}} [\href{http://inspirehep.net/record/1113003}{Inspire}].

\bibitem{Matsedonskyi:2012ws}
O.~Matsedonskyi, G.~Panico and A.~Wulzer, ``{Light Top Partners for a Light
  Composite Higgs}'', \href{http://dx.doi.org/10.1007/JHEP01(2013)164}{{\em
  JHEP} {\bfseries 1301} (2013) },
  \href{http://xxx.lanl.gov/abs/1204.6333}{{\tt arXiv:1204.6333}} [\href{http://inspirehep.net/record/1112825}{Inspire}].

\bibitem{Strumia:2011by}
A.~Strumia, ``{The fine-tuning price of the early LHC}'',
  \href{http://dx.doi.org/10.1007/JHEP04(2011)073}{{\em JHEP} {\bfseries 1104}
  (2011) }, \href{http://xxx.lanl.gov/abs/1101.2195}{{\tt arXiv:1101.2195}} [\href{http://inspirebeta.net/record/884116}{Inspire}].

\bibitem{Barbieri:1988p1552}
R.~Barbieri and G.~F. Giudice, ``Upper bounds on supersymmetric particle
  masses'', \href{http://dx.doi.org/10.1016/0550-3213(88)90171-X}{{\em Nucl.
  Phys.} {\bfseries B 306} (1988) 63} [\href{http://inspirebeta.net/record/248280}{Inspire}].

\bibitem{Giudice:961767}
G.~F. Giudice and R.~Rattazzi, ``{Living Dangerously with Low-Energy
  Supersymmetry. }'',
  \href{http://dx.doi.org/10.1016/j.nuclphysb.2006.07.031}{{\em Nucl. Phys.}
  {\bfseries B 757} (2006) 19--46},
  \href{http://xxx.lanl.gov/abs/hep-ph/0606105}{{\tt hep-ph/0606105}} [\href{http://inspirebeta.net/record/718897}{Inspire}].

\bibitem{ATLAS-CONF-2012-130}
{\bfseries ATLAS} Collaboration, ``{Search for exotic same-sign dilepton
  signatures (b' quark, $T_{5/3}$ and four top quarks production) in 4.7/fb of
  pp collisions at $\sqrt{s}=7$ TeV with the ATLAS detector}'', \href{http://cds.cern.ch/record/1478217}{ATLAS-CONF-2012-130}.

\bibitem{CMS-PAS-SUS-12-027}
{\bfseries CMS} Collaboration, ``{Search for RPV supersymmetry with three or
  more leptons and b-tags}'', \href{http://cds.cern.ch/record/1494689?ln=en}{CMS-PAS-SUS-12-027}.

\bibitem{DeSimone:2012ul}
A.~D. Simone, O.~Matsedonskyi, R.~Rattazzi and A.~Wulzer, ``{A First Top
  Partner's Hunter Guide}'', \href{http://dx.doi.org/10.1007/JHEP04(2013)004}{{\em JHEP} {\bfseries 1304}
  (2013)}, \href{http://xxx.lanl.gov/abs/1211.5663}{{\tt arXiv:1211.5663}} [\href{http://inspirehep.net/record/1203860}{Inspire}].

\bibitem{CMS-PAS-B2G-12-012}
{\bfseries CMS} Collaboration, ``{Search for T5/3 top partners in same-sign
  dilepton final state}'', \href{http://cds.cern.ch/record/1494689?ln=en}{CMS-PAS-B2G-12-012}.

\bibitem{Cirelli:2005br}
M.~Cirelli, N.~Fornengo and A.~Strumia, ``{Minimal Dark Matter}'',
  \href{http://dx.doi.org/10.1016/j.nuclphysb.2006.07.012}{{\em Nucl.Phys.}
  {\bfseries B753} (2005) 178--194},
  \href{http://xxx.lanl.gov/abs/hep-ph/0512090}{{\tt hep-ph/0512090}} [\href{http://inspirehep.net/record/699850}{Inspire}].

\bibitem{MatsedonskyiToAppear}
O.~Matsedonskyi, R.~Rattazzi, F.~Riva and T.~Vantalon, to appear.

\bibitem{Falkowski:2011ua}
A.~Falkowski, C.~Grojean, A.~Kaminska, S.~Pokorski and A.~Weiler, ``{If no
  Higgs then what?}'', \href{http://dx.doi.org/10.1007/JHEP11(2011)028}{{\em
  JHEP} (2011) }, \href{http://xxx.lanl.gov/abs/1108.1183}{{\tt
  arXiv:1108.1183}} [\href{http://inspirehep.net/record/922186}{Inspire}].

\bibitem{Bini:2011zb}
C.~Bini, R.~Contino and N.~Vignaroli, ``{Heavy-light decay topologies as a new
  strategy to discover a heavy gluon}'',
  \href{http://dx.doi.org/10.1007/JHEP01(2012)157}{{\em JHEP} {\bfseries 1201}
  (2012) }, \href{http://xxx.lanl.gov/abs/1110.6058}{{\tt arXiv:1110.6058}} [\href{http://inspirehep.net/record/943275}{Inspire}].

\bibitem{CLICToAppear}
R.~Contino, C.~Grojean, D.~Pappadopulo, R.~Rattazzi and A.~Thamm, to appear.

\bibitem{PappadopuloToAppear}
D.~Pappadopulo, A.~Thamm, R.~Torre and A.~Wulzer, to appear.

\bibitem{Falkowski:2013dza}
A.~Falkowski, F.~Riva and A.~Urbano, ``{Higgs At Last}'',
  \href{http://xxx.lanl.gov/abs/1303.1812}{{\tt arXiv:1303.1812}} [\href{http://inspirehep.net/record/1222861}{Inspire}].

\bibitem{Agashe:2002fu}
K.~Agashe, A.~Delgado and R.~Sundrum, ``{Gauge coupling renormalization in
  RS1}'', \href{http://dx.doi.org/10.1016/S0550-3213(02)00740-X}{{\em Nucl.
  Phys.} {\bfseries B643} (2002) 172--186},
  \href{http://xxx.lanl.gov/abs/hep-ph/0206099}{{\tt hep-ph/0206099}} [\href{http://inspirehep.net/record/588206}{Inspire}].

\end{thebibliography}\endgroup

\end{document}